\documentclass[conference]{IEEEtran}
\IEEEoverridecommandlockouts
\usepackage[T1]{fontenc}
\usepackage[utf8]{inputenc}
\usepackage[british,UKenglish,USenglish,american]{babel}
\usepackage{hyphenat}
\usepackage{hyperref}
\hypersetup{
    colorlinks=true,
    linkcolor=blue,
    filecolor=magenta,      
    urlcolor=black,
    citecolor=green,
}
\usepackage{cite}
\usepackage{gensymb}
\usepackage{amsmath,amssymb,amsfonts}
\usepackage[ruled]{algorithm2e}
\usepackage{graphicx}
\usepackage{subcaption}
\usepackage{textcomp}
\usepackage{xcolor}
\usepackage[noend]{algpseudocode}
\usepackage[flushleft]{threeparttable}

\def\BibTeX{{\rm B\kern-.05em{\sc i\kern-.025em b}\kern-.08em
    T\kern-.1667em\lower.7ex\hbox{E}\kern-.125emX}}
\begin{document}

\title{Data integration and prediction models of photovoltaic production from Brazilian northeastern\\
\thanks{This work is product from a partnership between Brazilian foudantations FITec, FACEPE and UPE university}
}

\author{
\IEEEauthorblockN{1\textsuperscript{st} Hugo Abreu Mendes}
\IEEEauthorblockA{\textit{POLI} \\
\textit{University of Pernambuco}\\
Recife, Brazil \\
ham@poli.br}

\and

\IEEEauthorblockN{2\textsuperscript{st} Henrique Ferreira Nunes}
\IEEEauthorblockA{\textit{FITec} \\
Recife, Brazil}

\and

\IEEEauthorblockN{3\textsuperscript{nd} Manoel da Nobrega Marinho}
\IEEEauthorblockA{\textit{POLI} \\
\textit{University of Pernambuco}\\
Recife, Brazil}

\and

\IEEEauthorblockN{4\textsuperscript{nd} Paulo S.G. de Mattos Neto}
\IEEEauthorblockA{\textit{CIN} \\
\textit{Federal University of Pernambuco}\\
Recife, Brazil}
}

\maketitle

\begin{abstract} 
 All productive branches of society need an estimate to be able to control their expenses well. In the energy business, electric utilities use this information to control the power flow in the grid. For better energy production estimation of photovoltaic systems, it is necessary to join multiples geospatial and meteorological variables. This work proposes the creation of a satellite data integration platform, with production estimation models, base stations measurement and actual production capacity. This work presents statistical, probabilistic and artificial intelligence models that generate spatial and temporal production estimates that could improve production gains as well as facilitate the monitoring and supervision of new enterprises are presented.
\end{abstract} 


\section{Introduction}
\label{sec:intro}
Estimating the potential of photovoltaic generation is a topic that has received good attention due to its importance and interest that the society has on the subject \cite{chin2015cell, jordehi2016parameter, de2017performance}. Previous works already takes into consideration the technology used by the cell and satellite models that aim to define the physical input parameters such as radiation, temperature and wind speed \cite{mueller2009cm, huld2012new, amillo2014new, habte2017evaluation}.

The use of AI (artificial intelligence) applied to this theme has been, in some way, limited to time series forecasting of generation \cite{voyant2017machine, wolff2016statistical, li2016hierarchical}. This type of prediction is very useful considering the complete electrical system of a region or country for balancing supply and demand, enabling greater predictability for the electrical system operator in relation to the appropriateness of choosing the correct and cleaner power sources at the right time. Because they require data from installed plants, these forecasting techniques are not used to evaluate new locations for photovoltaic projects.

Works that use AI techniques to predict photovoltaic generation even before the existence of the system is still incipient. These analyzes are made involving only GIS (Geographic Information System) data or specialized software \cite{khan2014optimal, fernandez2015site, carrion2008electricity, boran2010multi}.

This work aims to join the use of data science techniques and AI algorithms for spatial estimation of the photovoltaic generation potential, enabling a better choice of the location for the implantation of new photovoltaic plants. The work also includes the implementation of a hybrid model for time series forecasting, with daily average generation data from some plants.

The choice of northeast Brazilian region for case study takes into account the fact that it is the region with the largest amount of photo-voltaic generation data available through the ONS (Brazilian National System Operator) platform. \cite{ONS, ANEEL}.

The organization of the sections of this paper is as follows. In the section \ref{sec:espacial_nordeste} will be discussed spatial data obtained from different sources, all data in this section are obtained from time averages during all years of collection of each base. In sections \ref{sec:otmizacao_comite} and \ref{sec:cov-cor}, two models of spatial estimation of photovoltaic production are presented. Subsequently, section \ref{sec:series_temp_usinas} explains the use of time series for daily generation data in some northeastern plants. Finally, the results are presented in section \ref{sec:resultados}. Closing the paper, there is a brief discussion in section \ref{sec:discussao}, where improvements and ideas for posterity are raised.

\section{Spatial Data in the Brazilian Northeast}
\label{sec:espacial_nordeste}

The spatial distribution of the solar resource is a necessary knowledge, obtained from radioactive transfer models and validation with observational data. By combining data from the Brazilian Institute of Meteorology (INMET), PVGIS (Photovoltaic Geographic Information System), Brazilian ONS and the Brazilian Solar Energy Atlas of LABREN (Renewable Energy Resource Modeling and Studies Laboratory), it is possible to create a system capable of estimating the average capacity factor of a region, being the ratio between the generated production and the nominal production capacity. The capacity factor is used in these analyzes, as it is independent of the size of the plant and the technologies used in solar cells.

It can be integrated into the system of weather forecast intelligence, which can be daily or monthly generation for any plant. The complete system would therefore be able to provide a spatial estimate of production as well as a specific time estimate for an existing enterprise. The most important is, with the joining of databases, to facilitate the search for new ventures, for both utilities and microgenerators.

The following subsections describe the bases used for generating a data cube, where each column corresponds to a different variable. What enables the manipulation of this data is the fact, being the location (latitude and longitude).

\subsection{Brazilian Solar Energy Atlas}
This data comes from the Solarimetric Atlas, authored by LABREN. Data from 17 years (1999-2015) of collection, which results, finally, by averaging over time, in a radiation atlas in Brazil, with annual total and each month. \cite{pereira2017atlas, LABREN}. These data are obtained by satellite, using the BRASIL - SR model and validated with INMET stations \cite{martins2005base, martins2007mapeamento}. There are five columns in all about different ways of interaction of irradiation in the measurement environment.

\begin{itemize}
    
    \item {Global Horizontal} - Is the total energy per unit area incident on a horizontal surface, being the sum of the direct component and the diffuse component
    
    \item{Tilted} - Is the total energy per unit area incident on an inclined plane at the latitude of the location relative to the earth's surface
    
    \item{Direct Normal} - Is the total energy per unit of area coming directly from the sun that is perpendicular to the surface.
    
    \item{Diffuse} - It is the incident energy on a horizontal surface per unit area, resulting from the direct solar beam scattering by atmospheric constituents (molecules, particulate matter, clouds, etc.).
    
    \item{Photosynthetically active radiation} - Energy with photosynthetic production capacity, associated with biomass production, plant morphology and plant growth.
    
\end{itemize}

\subsection{INMET}
\begin{figure}[htbp]
    \includegraphics[width=1\columnwidth,trim={0.5 0.5 0.5 0.5},clip]{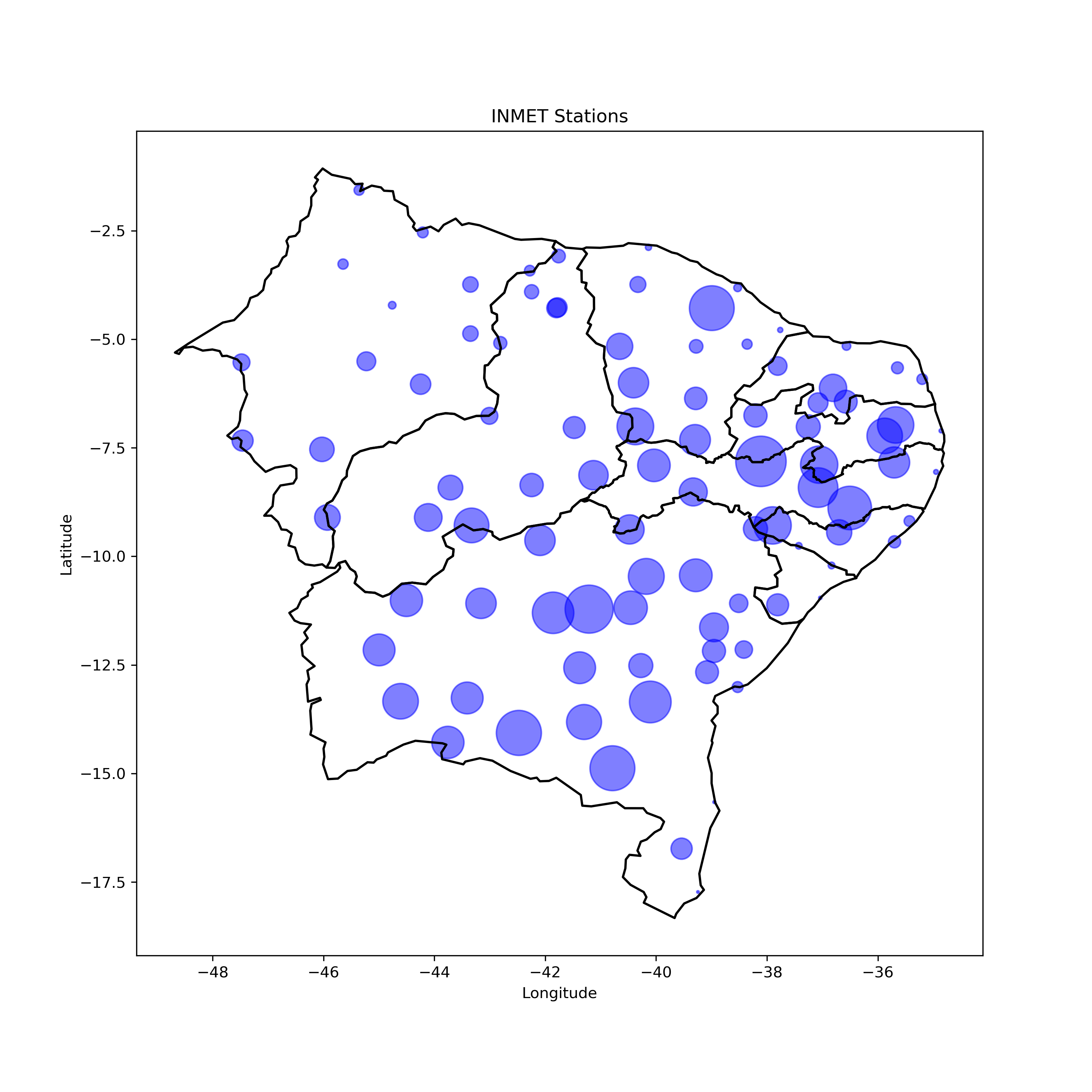}
    \caption{Location of INMET stations in northeastern Brazil. The larger the circle the higher the altitude.}
    \label{fig:inmet_locais}
\end{figure}

The National Institute of Meteorology provides data from various stations throughout Brazil \cite{INMET}. In this section these data are obtained for the same period as the data from the Brazilian Solar Energy Atlas, between 1999 and 2015, a temporal average is then made in these data, resulting only in their spatial component. In the northeast, one can see the location of the stations, by Figure \ref{fig:inmet_locais}. The following variables were used: 

\begin{enumerate}
    \item Total Solar Irradiance
    \item Number of Days with Precipitation
    \item Atmospheric pressure
    \item Average Maximum Temperature 
    \item Average Relative Humidity 
    \item Average Wind Speed
    \item Average Cloudiness
    \item Total Precipitation
    \item Average Compensated Temperature
    \item Average Visibility
    \item Average Minimum Temperature 
    \item Evaporation
\end{enumerate}

\begin{figure*}[htbp]
\begin{subfigure}{.5\textwidth}
  \centering
  \includegraphics[width = 0.85\columnwidth, trim={0.3 0.3 2.2cm 0.3},clip]{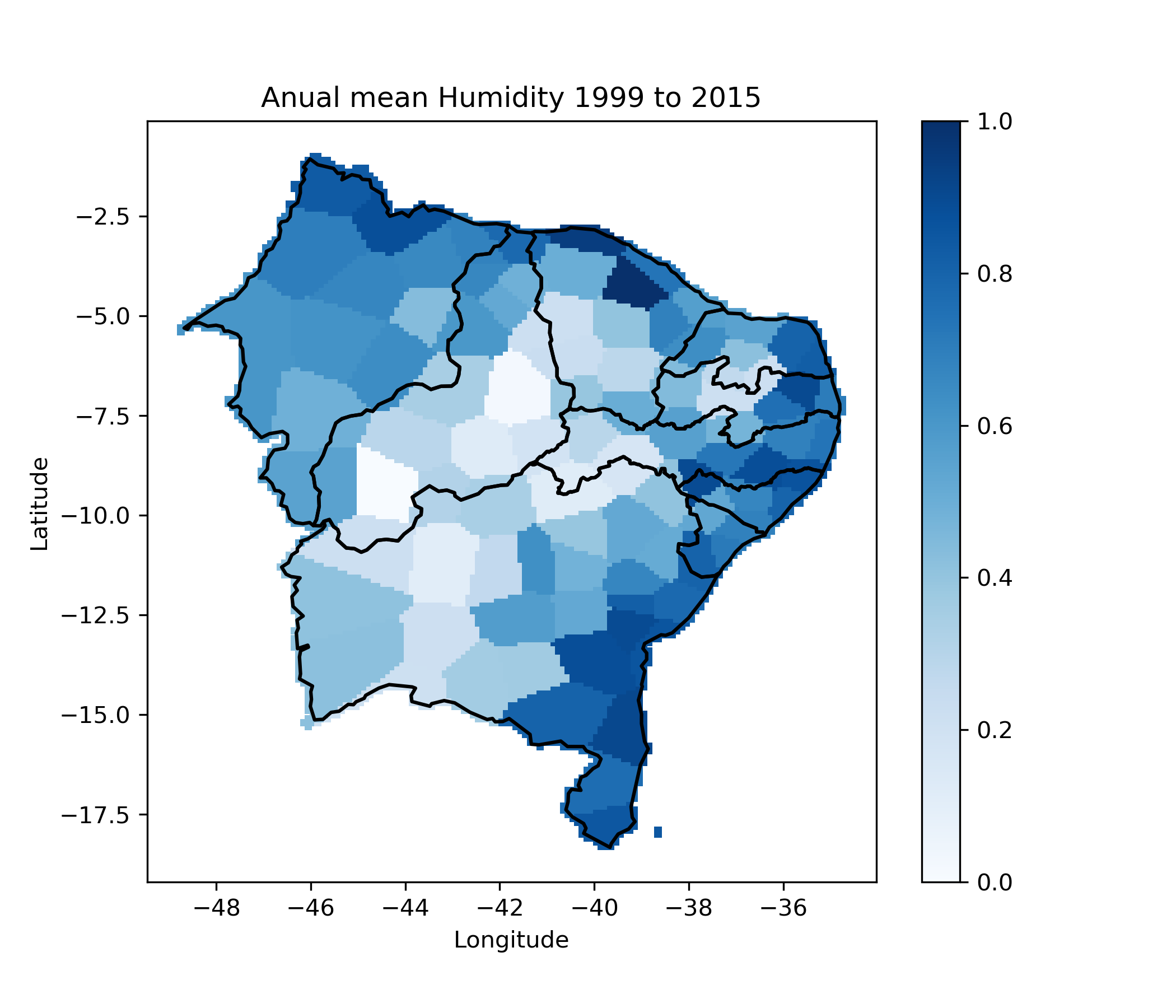}
  \caption{Average Relative Humidity}
  \label{fig:umidade}
\end{subfigure}%
\hfill
\begin{subfigure}{.5\textwidth}
  \centering
  \includegraphics[width = .85\columnwidth, trim={0.3 0.3 2.2cm 0.3},clip]{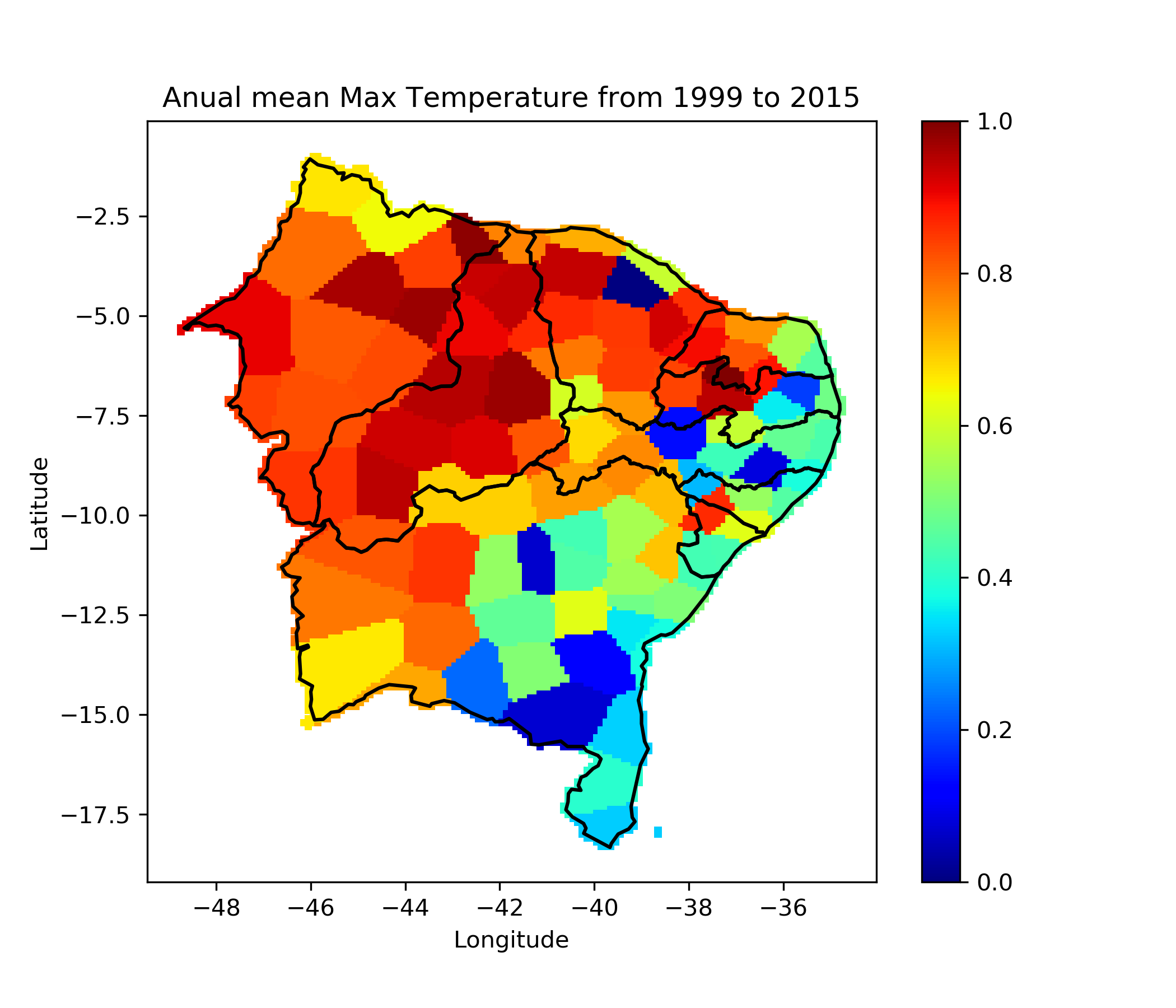}
  \caption{Average Maximum Temperature}
  \label{fig:tempmaximamedia}
\end{subfigure}
\begin{subfigure}{.5\textwidth}
  \centering
  \includegraphics[width = .85\columnwidth, trim={0.3 0.3 2.2cm 0.3},clip]{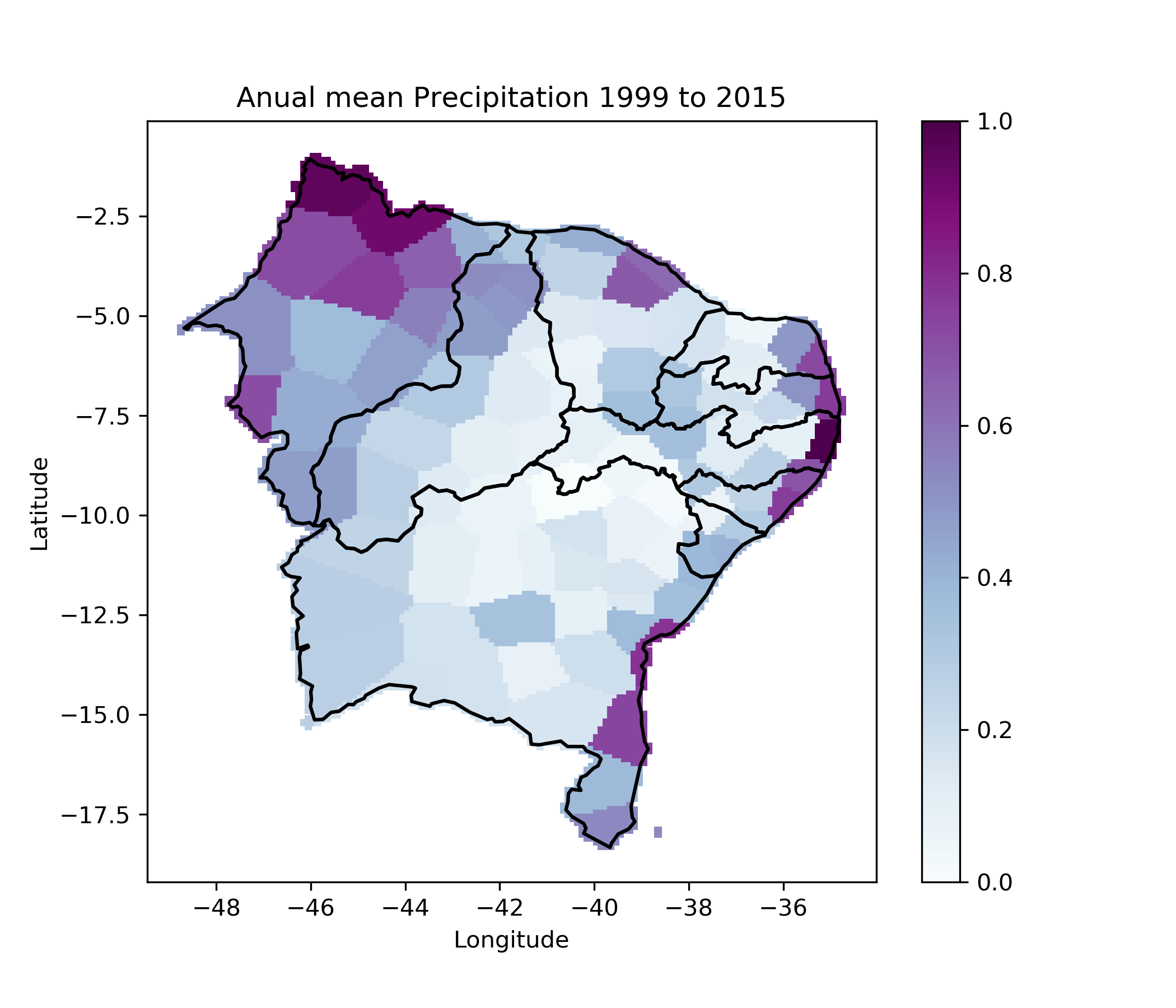}
  \caption{Total Precipitation}
  \label{fig:precipitacao}
\end{subfigure}%
\hfill
\begin{subfigure}{.5\textwidth}
  \centering
  \includegraphics[width = .85\columnwidth, trim={0.3 0.3 2.2cm 0.3},clip]{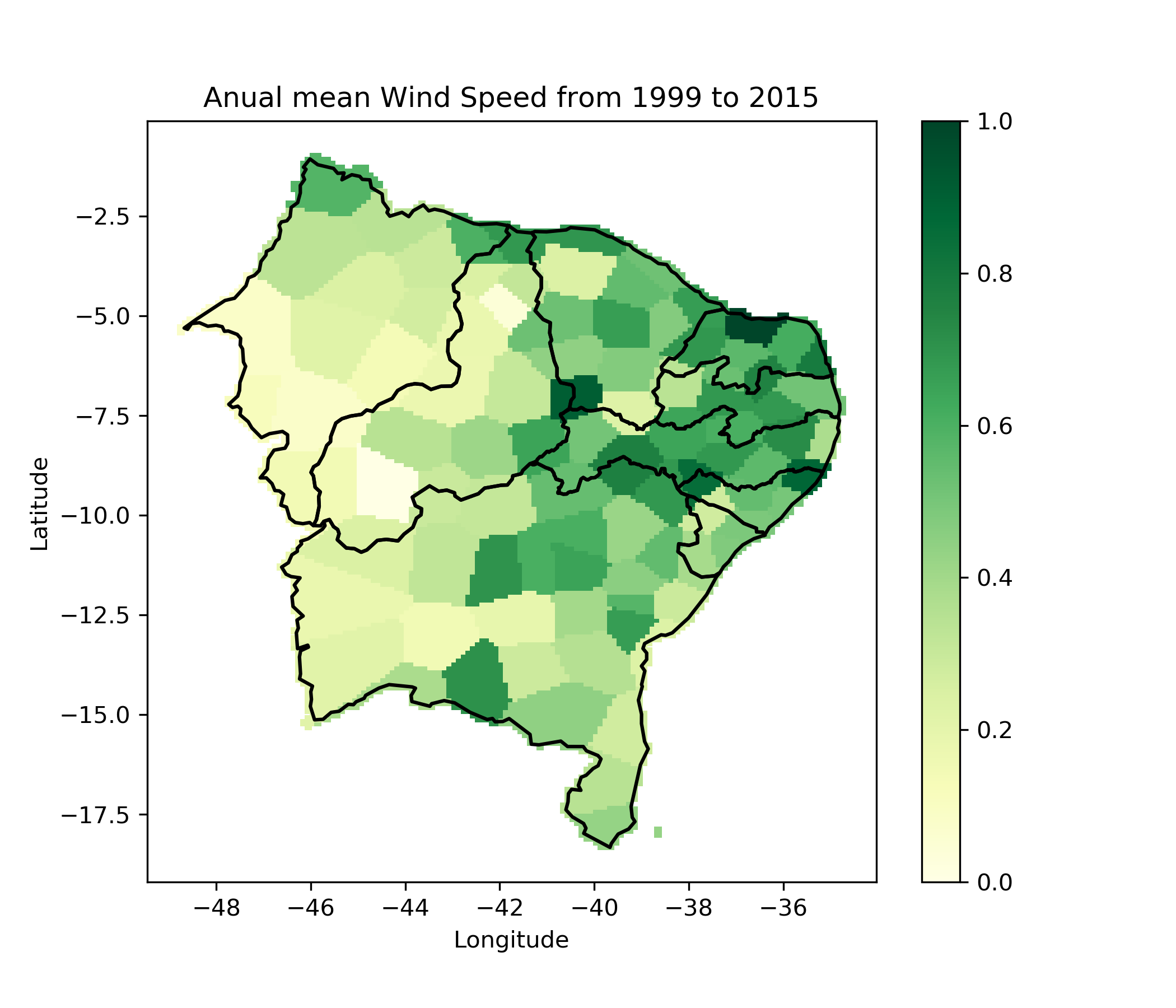}
  \caption{Average Wind Speed}
  \label{fig:velocidadevento}
\end{subfigure}
\begin{subfigure}{.5\textwidth}
  \centering
  \includegraphics[width = .85\columnwidth, trim={0.3 0.3 2.2cm 0.3},clip]{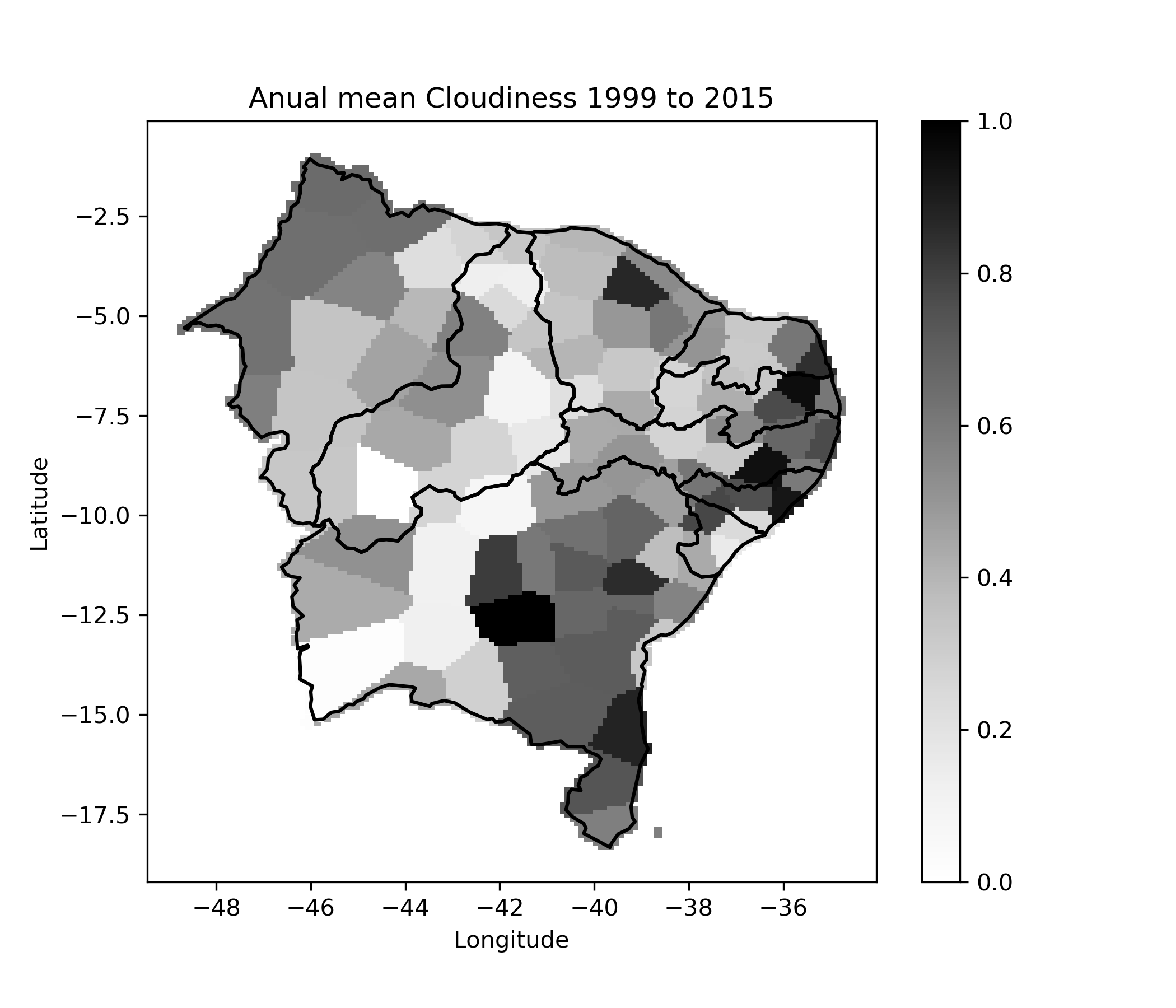}
  \caption{Average Cloudiness}
  \label{fig:nebulosidade}
\end{subfigure}%
\hfill
\begin{subfigure}{.5\textwidth}
  \centering
  \includegraphics[width = .85\columnwidth, trim={0.3 0.3 2.2cm 0.3},clip]{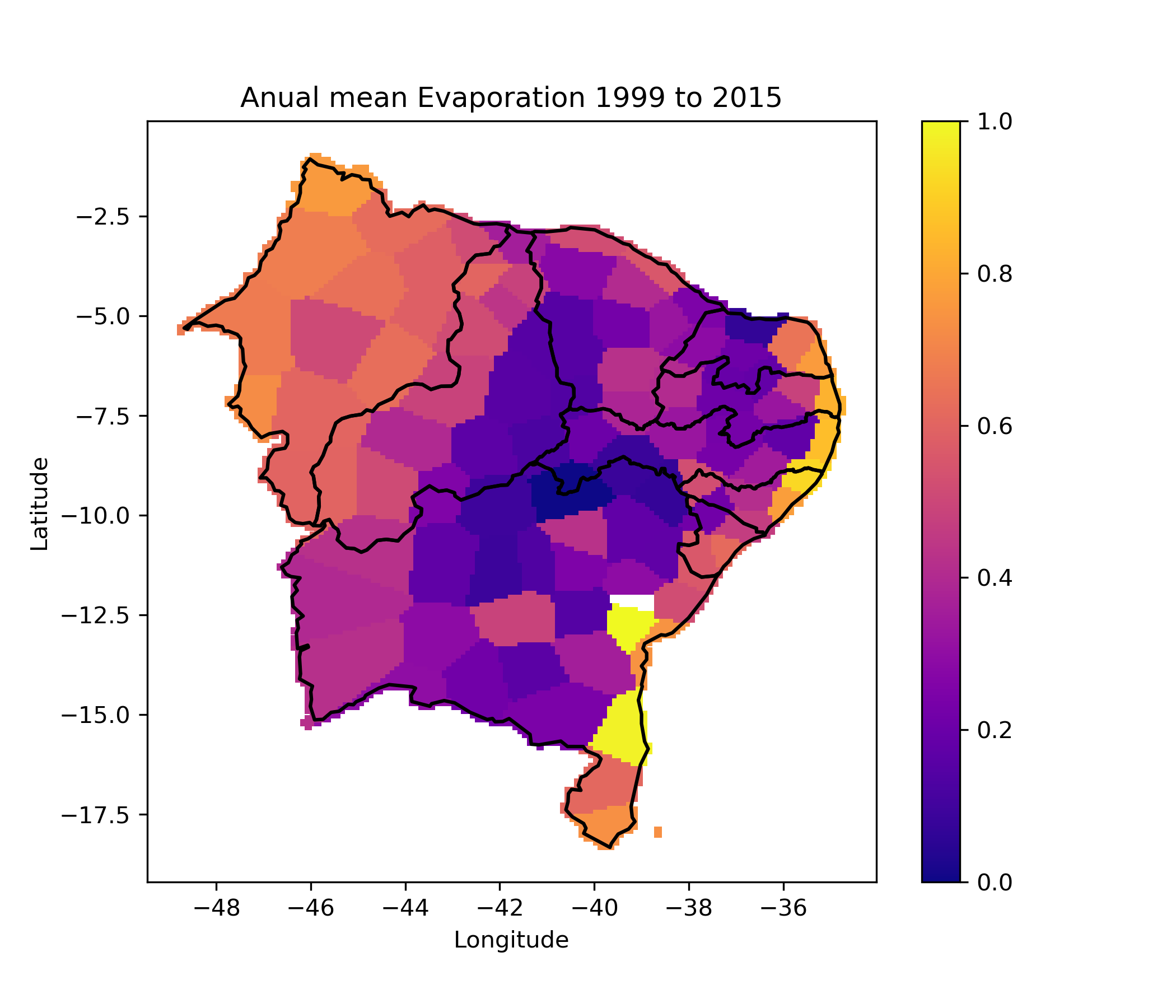}
  \caption{Evaporação}
  \label{fig:evaporacao}
\end{subfigure}
\caption{Compilation of normalized variables available from INMET}
\label{fig:INMET_data}
\end{figure*}

Some of these variables are compiled for viewing in Figure \ref{fig:INMET_data}, all variables were normalized to be scaled from zero to one.

\subsection{PVGIS}
It is a platform that has been developed for over 10 years by the European Commission JRC (Joint Research Center). PVGIS can be used to estimate the production of different types of photovoltaic systems virtually anywhere in the world \cite{PVGIS}. The PVGIS data is fully used, containing the following columns: monthly average (for all months of the year) of generation for the photovoltaic system chosen and standard deviation of monthly generation due to annual variations.

\begin{figure}[htbp]
    \includegraphics[width=1\columnwidth]{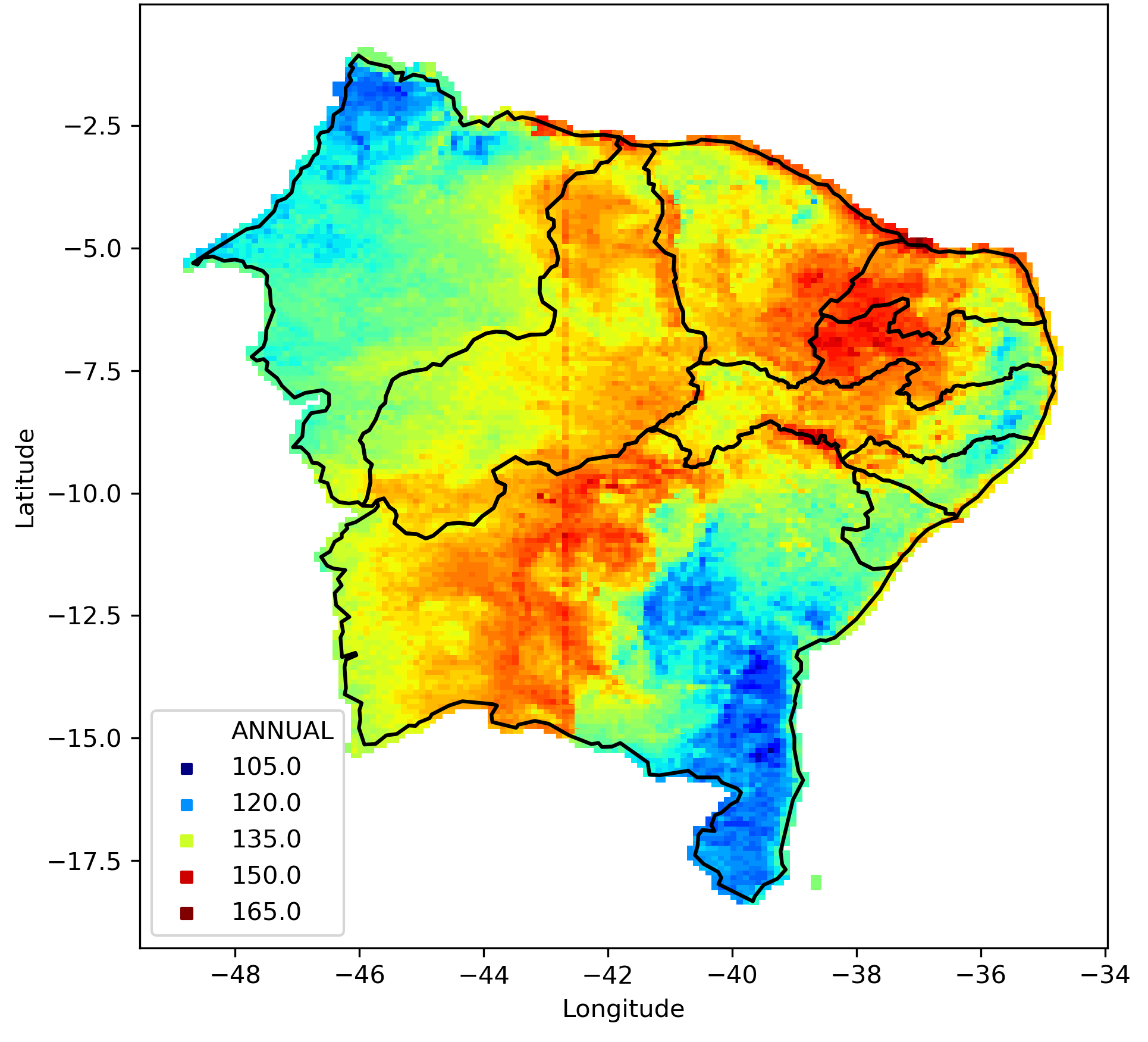}
    \caption{Monthly energy production map, obtained by modeled annual average, in kWh, from the NSRDB.}
    \label{fig:nsrdb_em}
\end{figure}

For the Brazilian northeast region, the PVGIS provides two satellite models, NSRDB \cite{habte2017evaluation} and SARAH \cite{muller2015sarah}. To obtain the PVGIS data, the web service was used, using 1kW peak power by default and optimum slope of the module \cite{pvgis_web}.

\begin{figure}[hbp]
    \includegraphics[width=1\columnwidth]{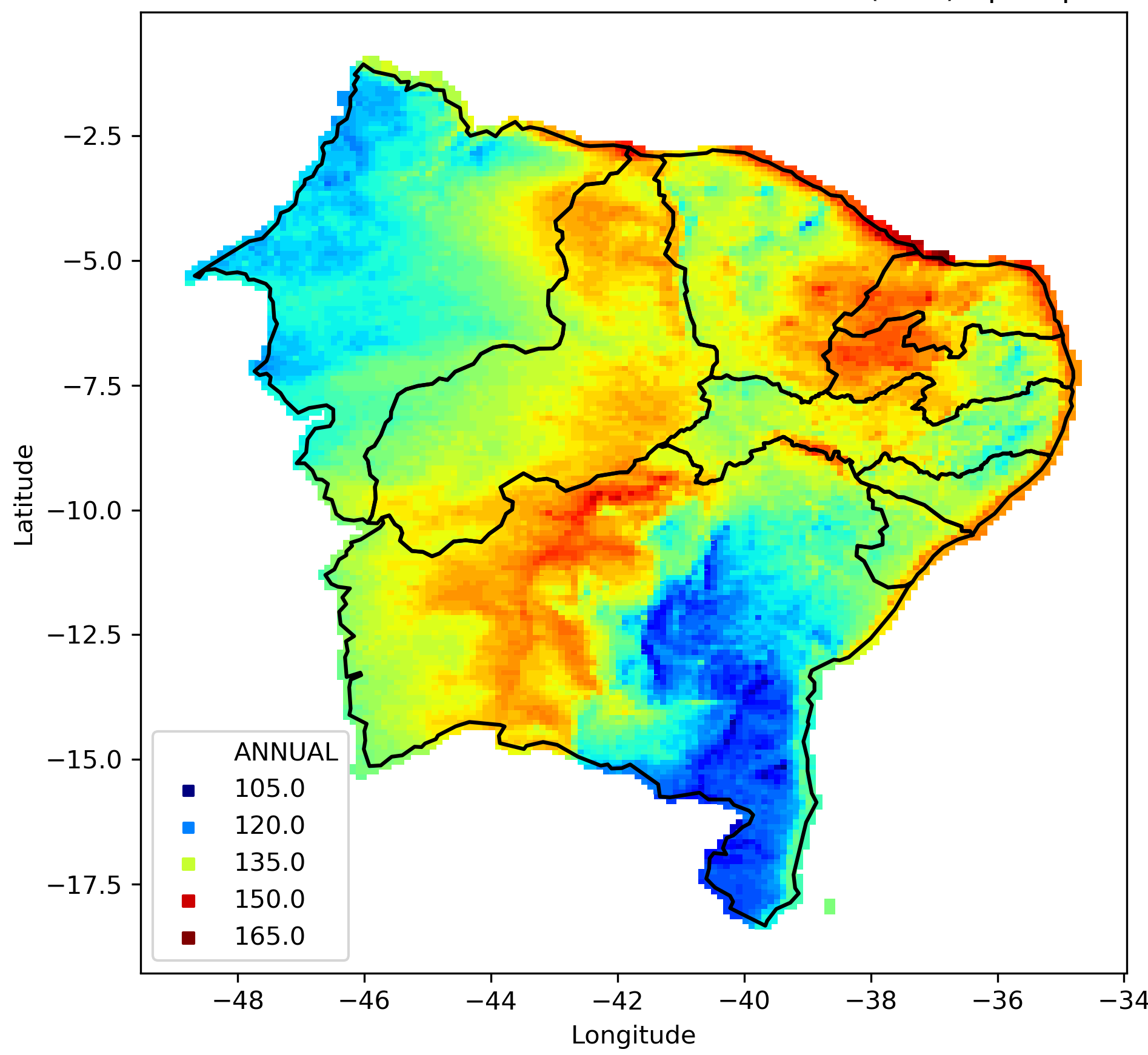}
    \caption{Monthly energy production map, obtained by the modeled annual average, in kWh, from the SARAH.}
    \label{fig:sarah_em}
\end{figure}

The estimated monthly average production by NSRDB and SARAH can be viewed respectively by Figures \ref{fig:nsrdb_em} e \ref{fig:sarah_em}. More about the methodologies and bases used by PVGIS can be found at \cite{PVGIS_methods, PVGIS_bases}

One reason for not using only PVGIS data to estimate energy production, is that it is not based on the BRASIL - SR radiation model, which is the most accurate as it is validated with local data \cite{martins2005base, martins2007mapeamento}. 

\subsection{ONS}
Of all the data already shown and discussed, none of them refer to actual generation information of photovoltaic systems. The problem is that actual data is available on a large scale only to installation companies and utilities. ONS makes available in its data balance of some plants in Brazil, but mostly from the Northeast.

\begin{figure}[htbp]
    \includegraphics[width=1\columnwidth,trim={1.5 1.5 2.0 2.5},clip]{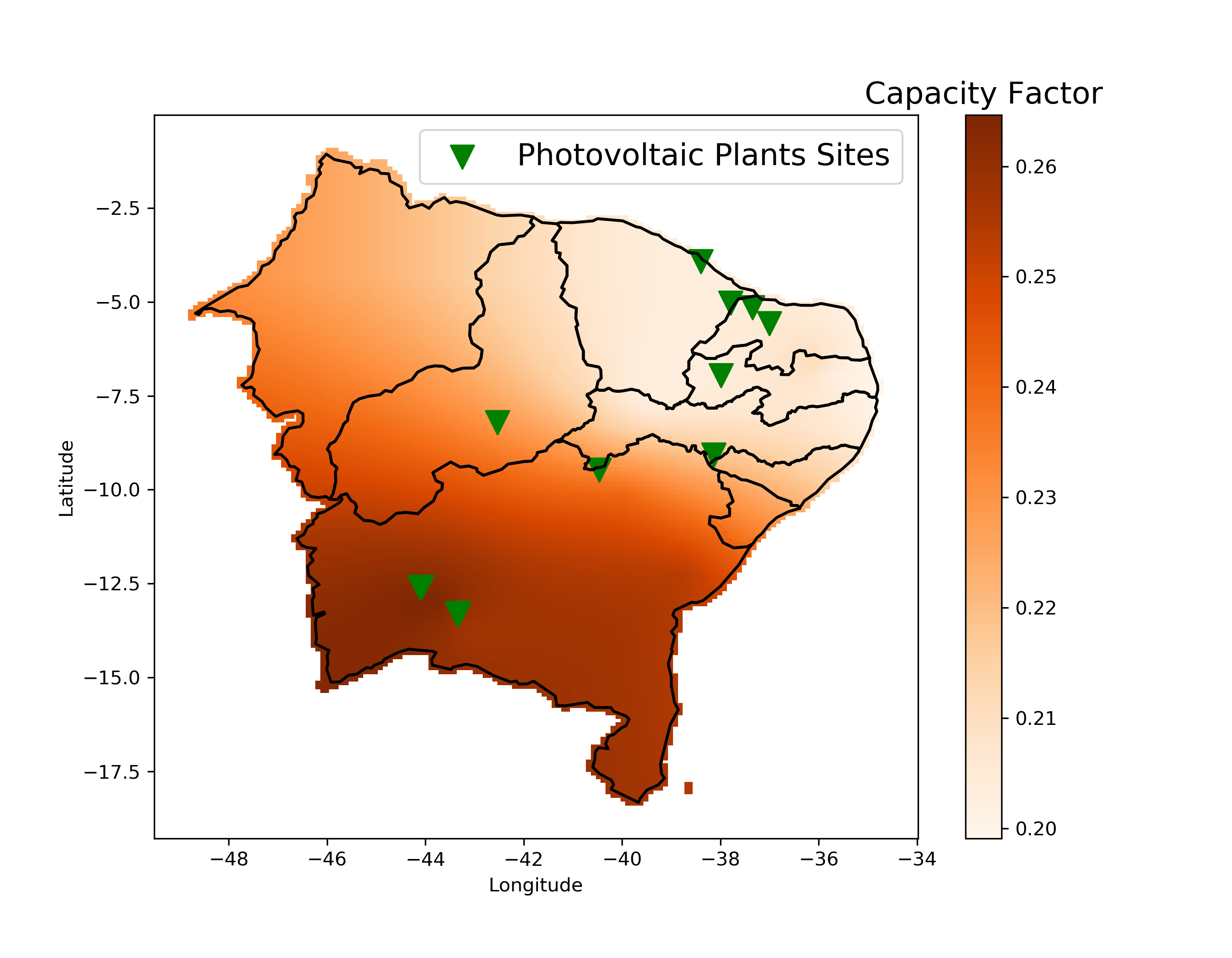}
    \caption{Map of generation capacity factor obtained by kriging interpolation and location of plants with data available by ONS in the Northeast}
    \label{fig:capacidade_IT}
\end{figure}

Having a more complete database containing power plants and off-grid generation (when the system is not connected to a network) data a capacity map could be better mapped. One methodology that can be used is Kriging interpolation \cite{stein2012interpolation, pykrige}. For our disposable data, the kriging results in Figure \ref{fig:capacidade_IT}, a variogram model based on power function was used \cite{oliver2014tutorial}. 

\section{Voting Regressor optimization}
\label{sec:otmizacao_comite}
From the data cube generated by concatenating all the spatial bases presented on the previous section, it is possible to create a regression model using ensemble voting intelligence. The intention is to visualize how AI would learn about the production capacity factor spatially, seen in Figure \ref{fig:capacidade_IT}. 

The vote used is by the average, the result is the mean of that obtained by several regression models: Linear regression; Random Forest Regressor \cite{svetnik2003random}; Support Vector Machine Regressor \cite{drucker1997support}; Adaboost \cite{solomatine2004adaboost}; Bagging \cite{breiman1996bagging}; GradientBoosting \cite{friedman2002stochastic}; RANSAC (Random Sample Consensus) \cite{fischler1981random}; Passive Agressive Regressor \cite{crammer2006online}; SGD (Stochastic Gradient Descendent) Regressor. All regressors were applied using the Python scikit-learn package \cite{scikit-learn}.

Each of the regressors has several different parameters, and not necessarily using them all is the best solution. For this, a search algorithm was made, aiming to make an optimal combination of regressors and their parameters.

The algorithm \footnote{Jupyter notebook:  \url{https://github.com/hugoabreu1002/Optm_ensenmble_data_cube/blob/master/Ensemble_on_capacidade.ipynb}} generates a big population, formed by combining these regressors with random parameters. Having formed this first random population, the algorithm makes cross-combinations between the regressors that are used in the committee with their respective parameters and selects those that best integrate the committee at each time.


\section{Production estimate based on Covariance and Correlation matrices}
\label{sec:cov-cor}

A case study\footnote{Jypter notebook:  \url{https://github.com/hugoabreu1002/PVIA-PE-CovCor/blob/master/Analise_CovCorr_KNN_e_Integrando_Bases.ipynb}} was done for Pernambuco, one of the nines states of brazilian northeast, from the data cube resulted of the methodology described in section \ref{sec:espacial_nordeste}. Seeking to generate a production estimate based only on covariance $\mathbf{K}$ and correlation $\mathbf{R}$ matrices of INMET and LABREN data variables.

The motivation for this estimate is to consider all possible variables, in addition to the direct radiation incident, in a given region. The challenge then is to define, which and how other variables will influence photovoltaic production.

First, all columns of the data cube must be normalized on a zero-unit scale. This is necessary so that the order of magnitude of each variable does not influence. After that, the covariance and correlation matrices of the data cube are necessary.

The estimate is given by $E$, on equation \ref{eq:1}. Where $\mathbf{X}$ is the column vector of all variables for each point in the discret space of Pernambuco's map; $\mathbf{a}$ is the row vector of the covariance between direct radiation and all others variables (INMET e LABREN), $\mathbf{b}$ is similarly, the line vector referring to the correlation.

\begin{equation}
E = \mathbf{a}\cdot{}\mathbf{X}\textsuperscript{2}+\mathbf{b}\cdot{}\mathbf{X}
\label{eq:1}
\end{equation}

The estimate consists of a linear combination of two terms, a quadratic and a linear. In the quadratic term, the variables of the column vector are potentiated and multiplied by the covariance, this indicates that, as all variables are less than or equal to the unit, the one that distances the most will influence less. In regions where the variables differ from their maximum, the effect of covariance information will be reduced.

The linear term, of the correlation, serves to insert the information of how much each variable influences the production, from its relation with the direct radiation. 

\begin{figure*}[htbp]
\begin{subfigure}{\textwidth}
  \centering
  \includegraphics[width = \textwidth]{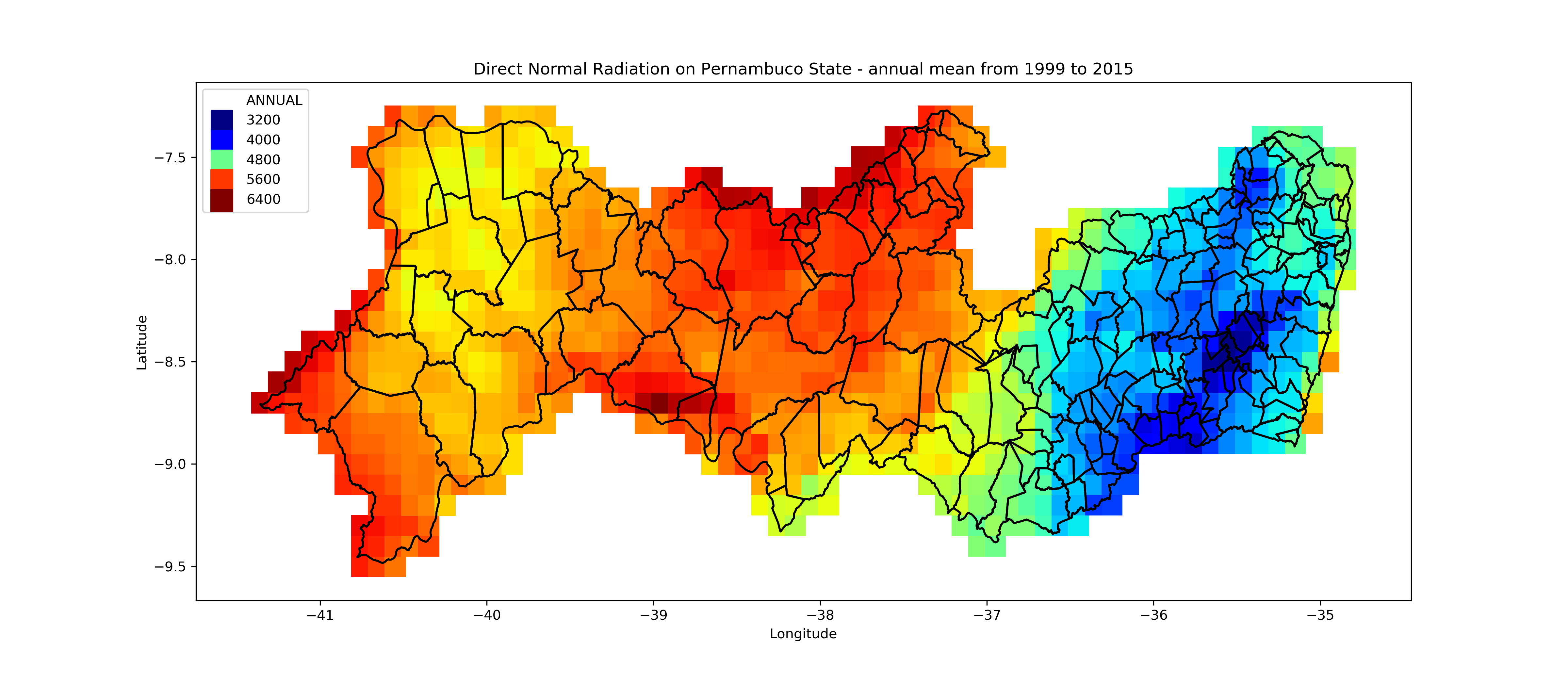}
  \caption{Normal direct radiation in the state of Pernambuco}
  \label{fig:radiacao_direta_pe}
\end{subfigure}
\begin{subfigure}{\textwidth}
  \centering
  \includegraphics[width = \textwidth]{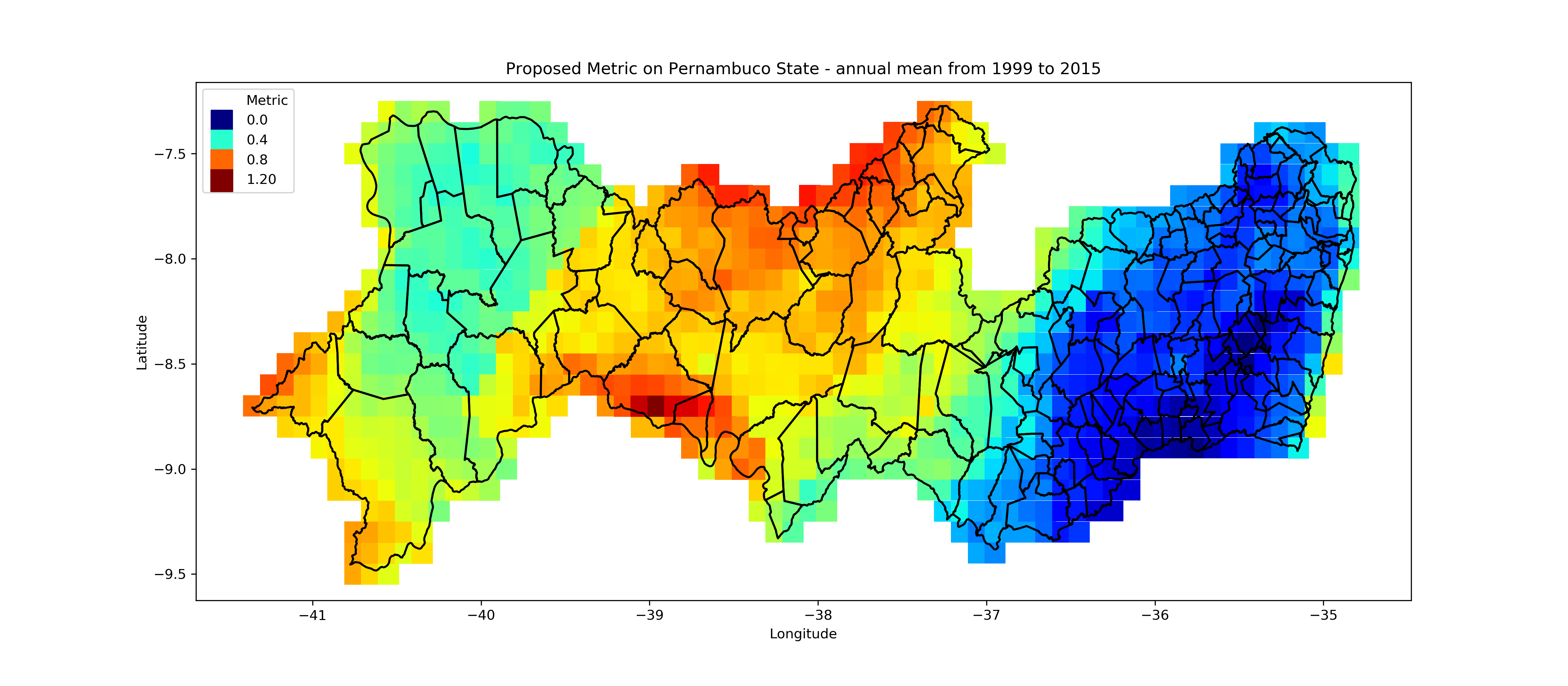}
  \caption{Proposed Metric}
  \label{fig:metrica_pe}
\end{subfigure}
\caption{Visualization of normal direct radiation in the state of Pernambuco and evaluation of the proposed metric}
\label{fig:pernambuco_metrica_rad}
\end{figure*}

The signs of the correlation values of the maximum mean temperature, humidity and precipitation variables were inverted, as they negatively influence the generation potential \cite{sun2017correlation, kazem2015effect, hailegnaw2015rain}. A visualization of the proposed estimation metric is possible with the Figure \ref{fig:pernambuco_metrica_rad}, along with the comparison with normal direct radiation.

\section{Data and time series models} 
\label{sec:series_temp_usinas}

The methodology used consists of generating a hybrid model between an ARIMA (Autoregressive Integrated Moving Average) and two ANN (Artificial Neural Networks), the ANNs used are MLP (Mult Layer Perceptron). First, the ARIMA model generates the forecast for the time series. The error, obtained by the difference between the time series and the ARIMA model, is saved in an error time series. In the second step, the first MLP is used to learn and forecast the error. Finally, the second MLP seeks the most suitable function to combine the ARIMA forecast with the error forecast. This type of modeling has been widely used and discussed in the literature \cite{zhang2003time, khashei2010artificial, babu2014moving, de2014hybrid, de2016hybrid, domingos2019intelligent}.

The time series used are obtained from the balance sheet provided by ONS \cite{ONS}. These are the average daily power generation of four photovoltaic plants. The ARIMA model used in this work makes use of exogenous variables, the literature references this use as an ARIMAX model. For one of the plants, seasonality information is also used, resulting in a SARIMAX model.

Exogenous variables are obtained from other time series from the INMET data \cite{INMET}, these exogenous variables are obtained for the same period as the energy generation time series. Exogenous variables were used: Precipitation, maximum temperature, minimum temperature, sunshine, average compensated temperature, average relative humidity and average wind speed. All series are manipulated after scaling between 0 and 1.

To obtain the two well-trained and parameterized MLPs, both for the non-linear association function and for error prediction, the entire model is surrounded by a genetic search algorithm. The algorithm looks for the best parameters for the MLP that estimates the error and the MLP that generates a non-linear function that associates pure ARIMA and modeled error.

In this search, the algorithm generates a population of MLPs, with random parameters, evaluates these and ranks the best parameters. After that, a new population is generated, after a cross between better and worse MLPs. The best MLP is repeated in the next generation. After crossing, the numerical parameters change. This new population is re-evaluated and the cycle continues.

For all series, 80\% of training data was stipulated, which is used to perform the training of the optimization and hybridization algorithm, as well as all MLPs and 20\% for testing, which is used to evaluate the final hybrid model.

On top of that, another evolutionary search is carried out to find, as variables of \textit {Lag}, amount of data from the past that is used to visualize the next future data and \textit {Forecast}, amount of forecasts from a model. The variables in question are described:

\begin{itemize} 
    
    \item \textit{Lag Error}: Number of samples from the past used to predict the next future sample of the time series of the error, obtained from the point-to-point subtraction between the original time series and the one obtained by the ARIMA model.
    
    \item \textit{Forecast Association Error} Number of future samples generated by the modeled error. It is used as one of the inputs of the associative function between modeled error and ARIMA series.
    
    \item\textit{Lag Association Error} Number of samples from the past of the modeled error, is also used as input to the associative function between modeled error and ARIMA series.
    
    \item \textit{Lag Association ARIMA} Number of samples from the past of the ARIMA model series: obtained over the ARIMA series, it is used as an input for the associative function between modeled error and ARIMA series.
    
\end{itemize}

The chosen genetic algorithm does not take into account the implicit variability of MLP training, however it was programmed in order to save the models with the trained weights, in each generation, not just the topologies.

Each MLP used in the proposed hybrid model has characteristics as shown below:
\begin{itemize}
    \item Activation function \{identity, logistics, hyperbolic tangent, relu\}
    \item Learning Rate Update \{constant, invscaling, adaptive\}
    \item Solver: adam, lbfgs.
    \item Hidden Layers Topology.
\end{itemize}

\section{Results}
\label{sec:resultados}

\subsection{Voting Regressor Optimization \ref{sec:otmizacao_comite}}

\begin{figure*}[htbp]
\begin{subfigure}{.45\textwidth}
  \centering
  \includegraphics[width=.9\textwidth, trim={0.1 0.5 1.5 1.0}]{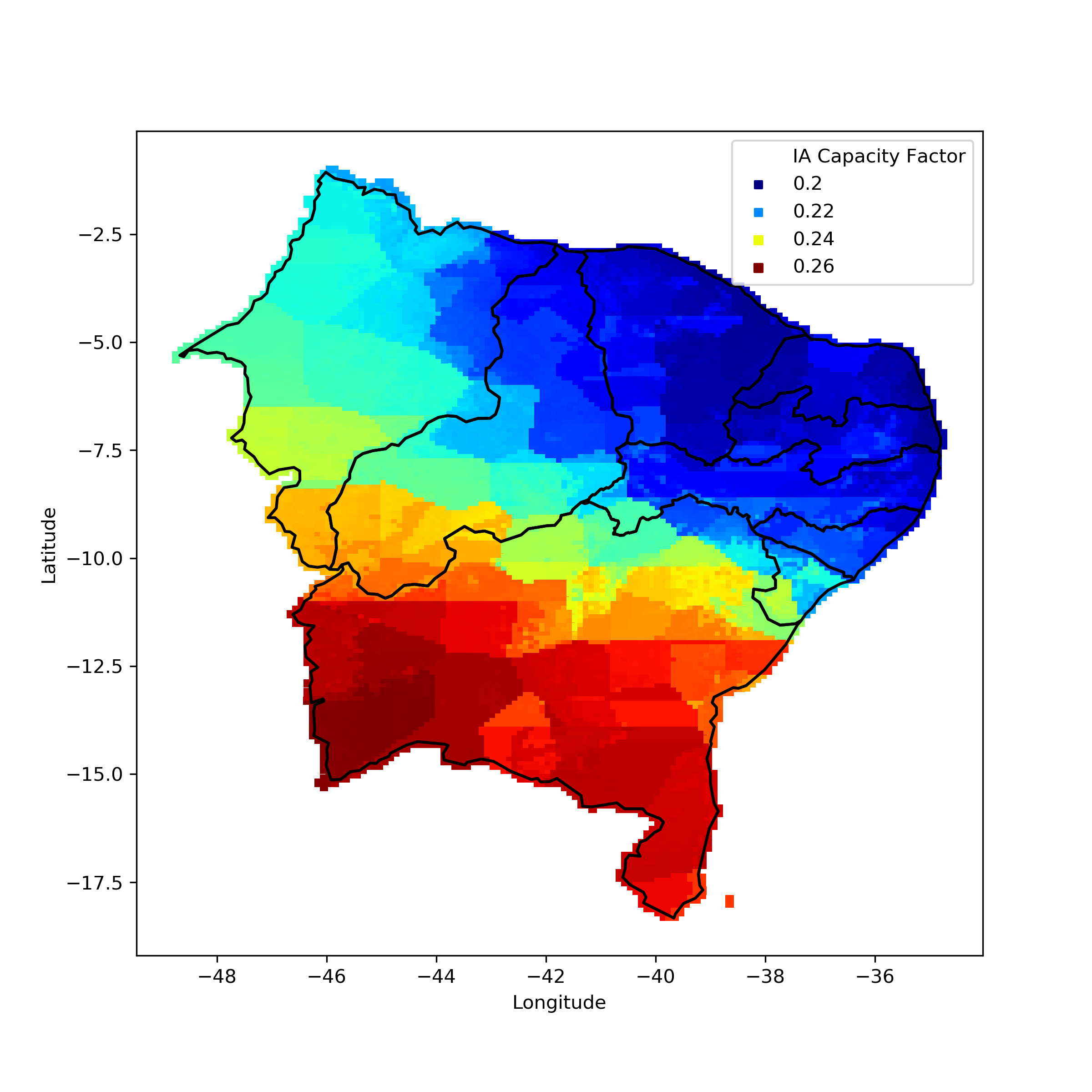}
  \caption{Capacity factor obtained from the optimized voting regressor}
  \label{fig:capacidade_IA}
\end{subfigure}%
\hfill
\begin{subfigure}{.45\textwidth}
  \centering
  \includegraphics[width=.9\textwidth, trim={0.1 0.5 1.5 1.0}]{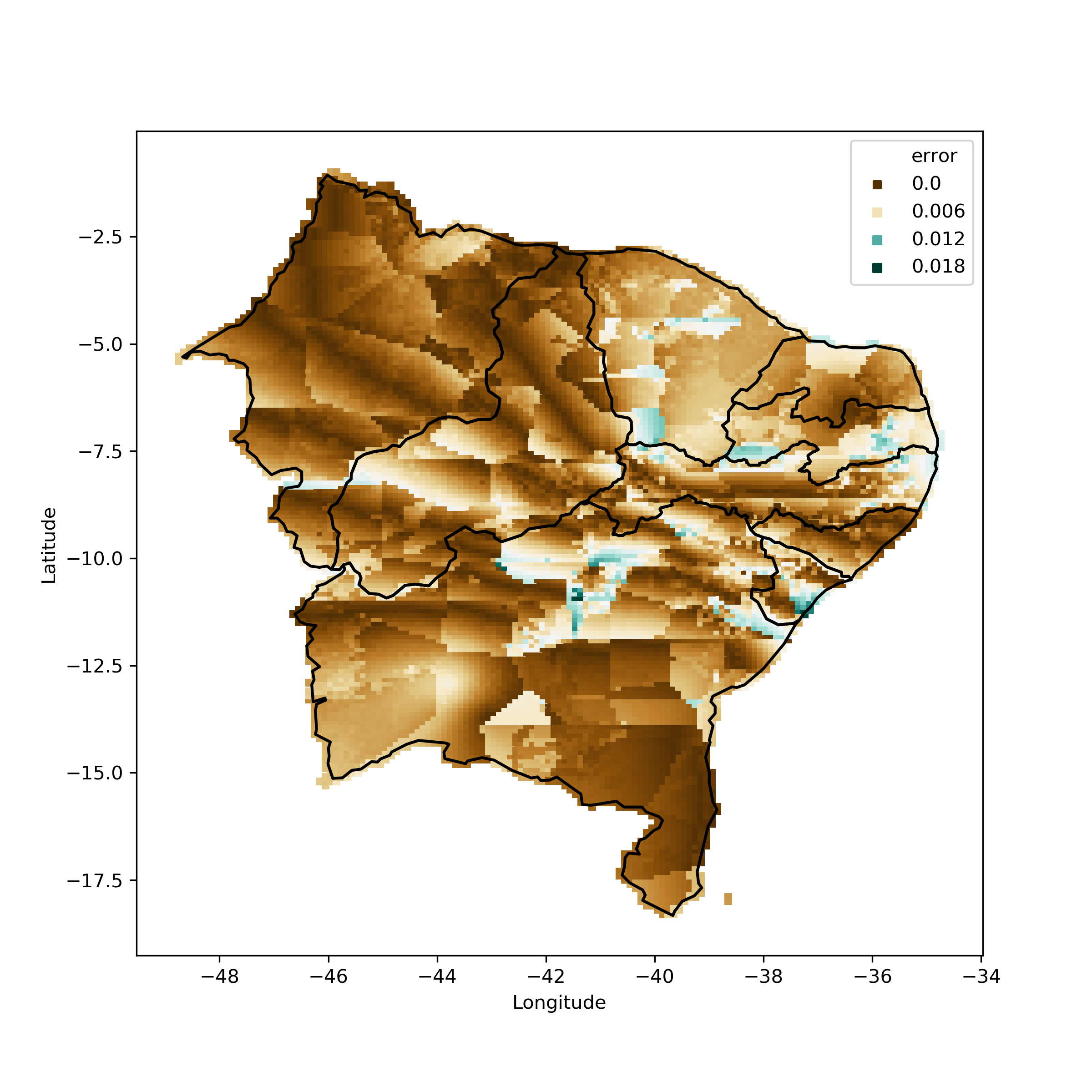}
  \caption{Error between AI model capacity factor and interpolated capacity factor of Figure\ref{fig:capacidade_IT}}
  \label{fig:erro_capacidade_IA}
\end{subfigure}
\caption{Comparison between interpolated spatial capacity factor and capacity factor obtained by AI model}
\label{fig:resultado_otm_comite}
\end{figure*}

The section \ref{sec:otmizacao_comite} presents a methodology for the optimization of voting regressors. The result leads to a reduction of the error between the actual interpolated production capacity of Figure \ref{fig:capacidade_IT} and the capacity given by the optimized voting regressor. When comparing a non optimized voting regressor, when all objects (the regressors participating of the voting) are initialized with the default configuration of python scikit-learn package \cite{scikit-learn}, there is a MAE reduction of 10,43\%, (23\% MSE).

The Figure \ref{fig:capacidade_IA} is the visual result of the optimized voting regressor, while the Figure \ref{fig:erro_capacidade_IA} show the error in o error with respect to the interpolated capacity factor of Figure \ref{fig:capacidade_IT}. It is seen from these Figures, and looking back to Figure \ref{fig:INMET_data} that the AI has some difficulty in smoothing the boundaries of the INMET stations.

\subsection{Production estimate based on Covariance and Correlation matrices \ref{sec:cov-cor}}

Based on the methodology described in the section \ref{sec:cov-cor}, it is possible to assess whether the proposed estimate is closer to the actual photovoltaic potential than normal direct radiation. For that, it is necessary that all data are on the same scale and make a comparison with metrics MSE (Mean Squared Error) and MAE (Mean Absolute Error), with the satellite model data PVGIS NSRDB and SARAH, described in the section \ref{sec:espacial_nordeste}. This can be assessed from the Table \ref{tab:metrica_pe}.

\begin{table}[htbp]
\caption{Error assessment between satellite models, normal direct radiation and proposed metric.}
\begin{center}
\begin{tabular}{ccccc}
  &\multicolumn{2}{c}{NSRDB}&\multicolumn{2}{c}{Sarah}\\\hline
  &MSE&MAE&MSE&MAE \\ \hline
Radiação Direta Normal&0.0203& 0.1247& 0.0515& 0.1993\\\hline
Métrica Proposta&\textbf{0.0194}&\textbf{0.1057}&\textbf{0.0263}&\textbf{0.1310}\\\hline
\label{tab:metrica_pe}
\end{tabular}
\end{center}
\end{table}

\subsection{Hybrid time series forecast model \ref{sec:series_temp_usinas}}

\begin{figure*}[htbp]
\begin{subfigure}{.45\textwidth}
  \centering
  \includegraphics[width=.9\textwidth]{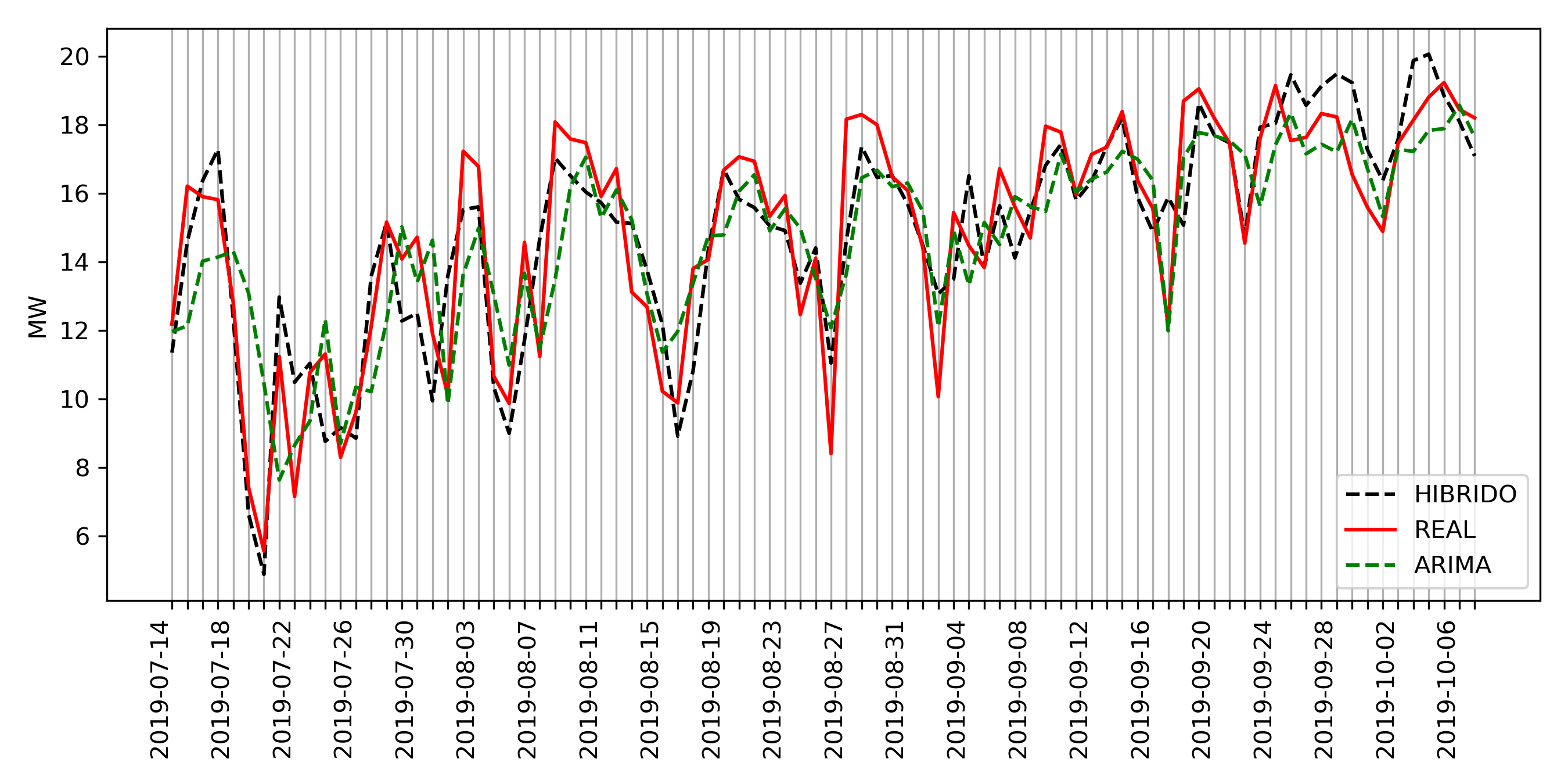}
  \caption{Rio Alto photovoltaic plant, located in Coremas-PB}
  \label{fig:rio_alto}
\end{subfigure}%
\hfill
\begin{subfigure}{.45\textwidth}
  \centering
  \includegraphics[width=.9\textwidth]{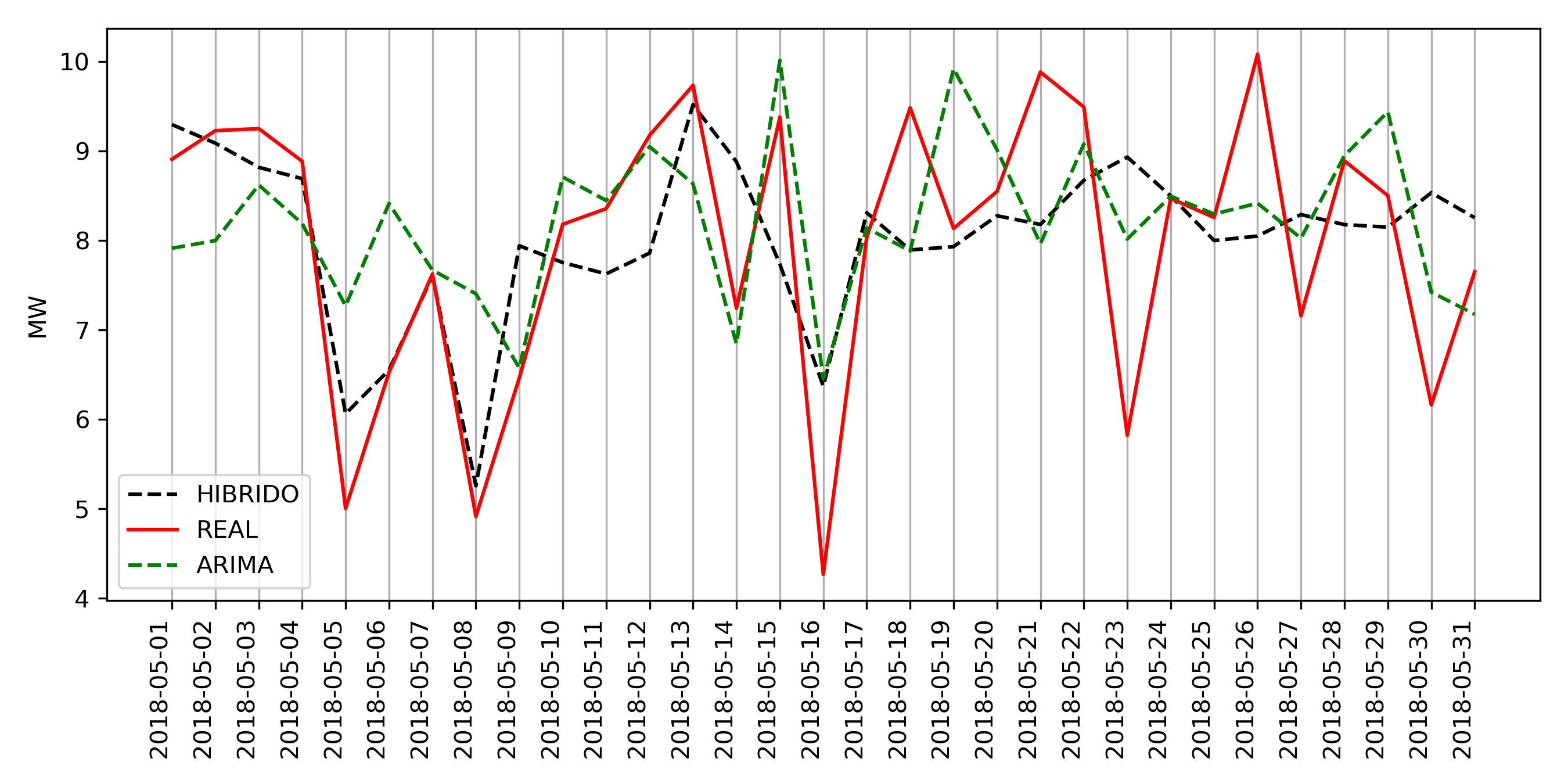}
  \caption{Photovoltaic plant Assu 5, located in Assu-RN}
  \label{fig:assu_5}
\end{subfigure}
\begin{subfigure}{.45\textwidth}
  \centering
  \includegraphics[width=.9\textwidth]{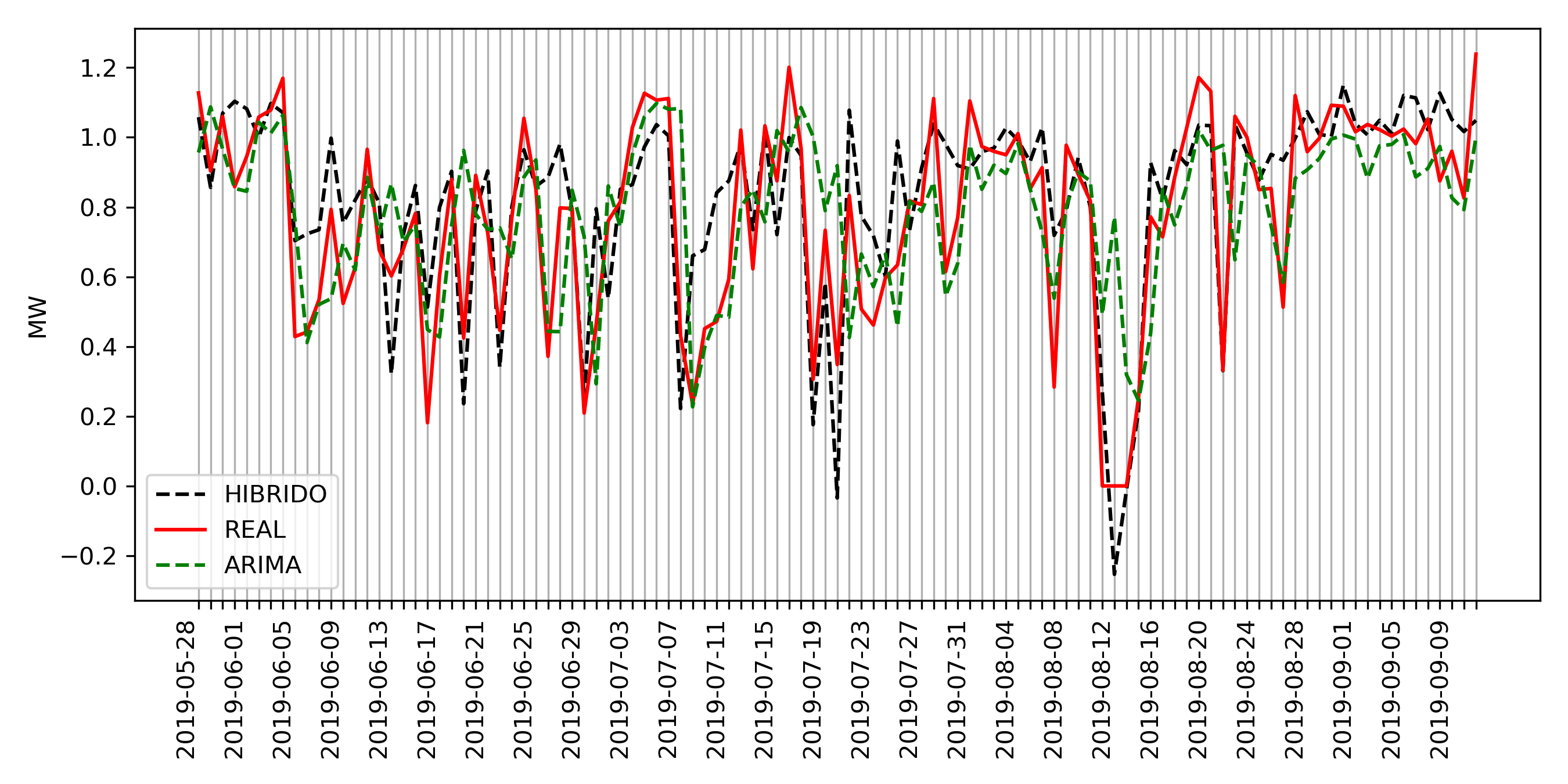}
  \caption{Fontes Solar 1 photovoltaic plant, located in Tacaratu-PE}
  \label{fig:fontes_solar_1}
\end{subfigure}%
\hfill
\begin{subfigure}{.45\textwidth}
  \centering
  \includegraphics[width=.9\textwidth]{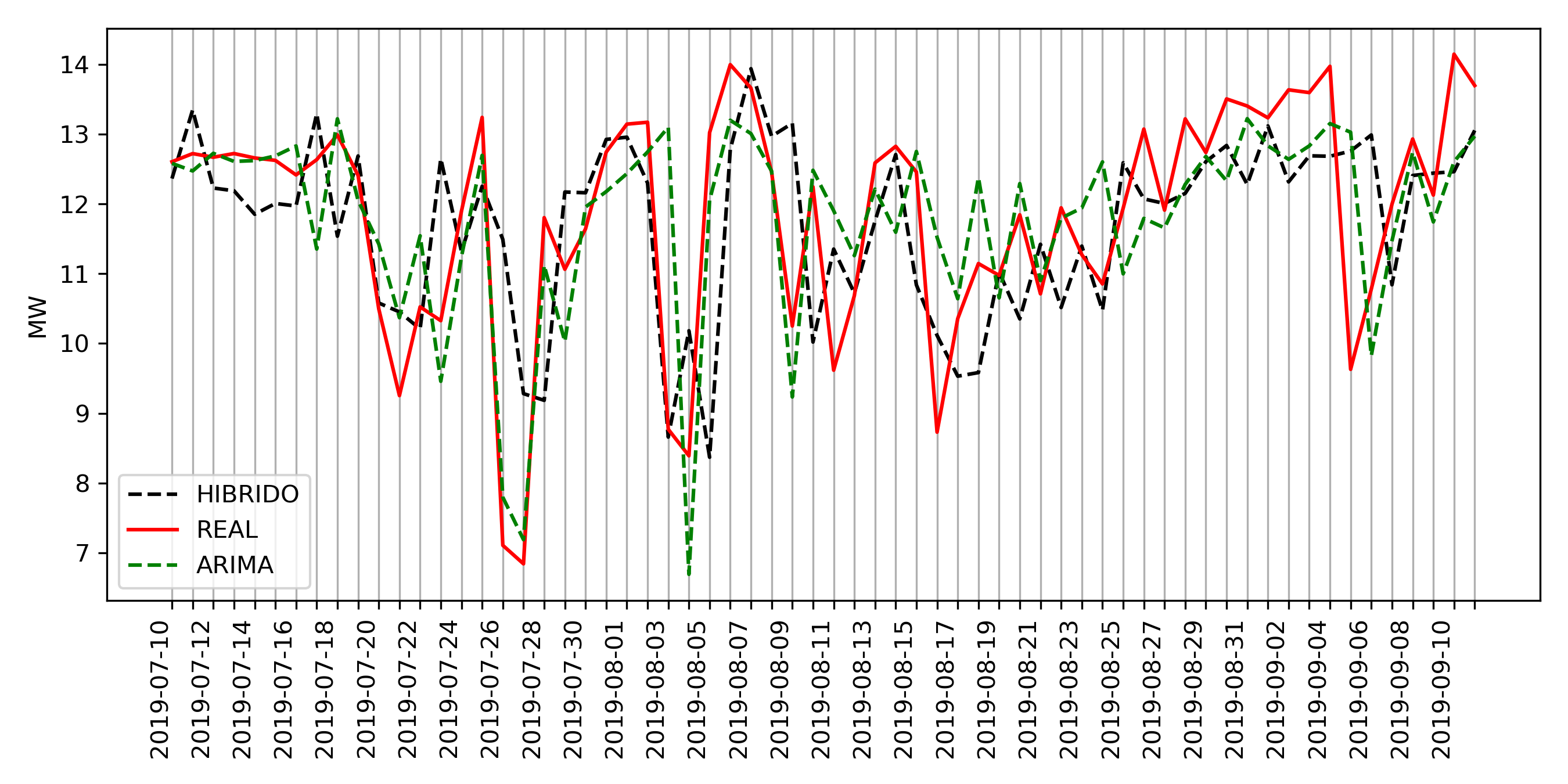}
  \caption{BJL Solar photovoltaic plant, located in Bom Jesus da Lapa-BA}
  \label{fig:BJL_solar}
\end{subfigure}
\caption{Time series of daily average generation, hybrid model and ARIMA model of some plants in the Northeast.}
\label{fig:resultado_time_series}
\end{figure*}

Based on the methodology described in the section \ref{sec:series_temp_usinas}, the result of the hybrid model for some plants is shown in Figure \ref{fig:resultado_time_series}. Numerically taking the Table \ref{tab:resultado_time_series}, which uses some metrics to evaluate the same result only on the test data, these metrics are used on the unit scale data.

Each of the daily average generation series resulted in an optimized topology for the hybrid model. These topologies are shown below. All results can be accessed \footnote{\url{https://github.com/hugoabreu1002/series_temp_hibrid_ons}}.

\begin{table}[htbp]
\caption{Evaluation of the result of the time series prediction algorithm using MAE, MSE and MAPE (Mean Absolute Percentage Error) metrics}
\begin{center}
\label{tab:resultado_time_series}
\begin{tabular}{cccl|ccl}
\hline
\textbf{} & \multicolumn{3}{c|}{Rio Alto} & \multicolumn{3}{c|}{Assu 5} \\ \hline
\textbf{} & MAE & MSE & MAPE & MAE & MSE & MAPE \\ \hline
ARIMA & 0.0722 & 0.0089 & 0.1084 & 0.0888 & 0.0132 & 0.1370 \\ \hline
Híbrido & \textbf{0.0640} & \textbf{0.0067} & \textbf{0.0923} & \textbf{0.0839} & \textbf{0.0126} & \textbf{0.1250} \\ \hline
\multicolumn{1}{l}{} & \multicolumn{3}{c|}{Fontes Solar 1} & \multicolumn{3}{c|}{BJL Solar} \\ \hline
\multicolumn{1}{l}{ARIMA} & \multicolumn{1}{l}{0.1181} & \multicolumn{1}{l}{0.0285} & 0.2647 & \multicolumn{1}{l}{\textbf{0.0466}} & \multicolumn{1}{l}{\textbf{0.0042}} & \textbf{0.0728} \\ \hline
\multicolumn{1}{l}{Hibrido} & \multicolumn{1}{l}{\textbf{0.1049}} & \multicolumn{1}{l}{\textbf{0.0184}} & \textbf{0.2511} & \multicolumn{1}{l}{0.0626} & \multicolumn{1}{l}{0.0072} & 0.0975
\end{tabular}
\end{center}
\end{table}

\subsubsection{Rio Alto}
In this series, the last 86 points are used as test data. Being the first 340 for training. This result was obtained from 100 individuals and 7 training periods of the evolutionary algorithm used. For this series the final models found by the algorithms are described below:
\begin{itemize}
\item ARIMA: P=1, D=1, Q=1; 
\item MLP For Error Modeling
    \begin{itemize}
        \item activation='identity'; 
        \item learning rate = 'adaptive';
        \item solver='lbfgs';
        \item hidden layers sizes = (114, 30, 10); 
        \item \textit{Lag Error} = 10.
    \end{itemize}
\item MLP For Nonlinear Association Function
    \begin{itemize}
        \item activation = 'identity'; 
        \item learning rate = 'invscaling';
        \item solver = 'lbfgs'; 
        \item hidden layers sizes = (35, 44, 3); 
        \item \textit{Lag  Association ARIMA} = 13; 
        \item \textit{Lag Association Error} = 13; 
        \item \textit{Forecast Association Error} = 15.
    \end{itemize}
\end{itemize}

\subsubsection{Assu 5}
In this series, the last 31 points are used as test data. The first 122 for training. This result was obtained from 100 individuals and 7 training periods of the evolutionary algorithm used. For this series the final models found by the algorithms are described below:
\begin{itemize}
\item ARIMA: P=0, D=1, Q=1; 
\item MLP For Error Modeling
    \begin{itemize}
        \item activation='identity'; 
        \item learning rate = 'invscaling';
        \item solver='adam';
        \item hidden layers sizes = (78, 51, 8); 
        \item \textit{Lag Error} = 6.
    \end{itemize}
\item MLP For Nonlinear Association Function
    \begin{itemize}
        \item activation = 'tanh'; 
        \item learning rate = 'constant';
        \item solver = 'lbfgs'; 
        \item hidden layers sizes = (65, 30, 3); 
        \item \textit{Lag  Association ARIMA} = 7; 
        \item \textit{Lag Association Error} = 2; 
        \item \textit{Forecast Association Error} = 15.
    \end{itemize}
\end{itemize}

\subsubsection{Fontes Solar 1}
In this series, the last 107 points are used as test data. The first 428 for training. This result was obtained from 100 individuals and 7 training periods of the evolutionary algorithm used. For this series the final models found by the algorithms are described below:
\begin{itemize}
\item ARIMA: P=1, D=1, Q=1;
\item MLP For Error Modeling
    \begin{itemize}
        \item activation='tanh'; 
        \item learning rate = 'adaptive';
        \item solver='adam';
        \item hidden layers sizes = (9, 30, 7); 
        \item \textit{Lag Error} = 4.
    \end{itemize}
\item MLP For Nonlinear Association Function
    \begin{itemize}
        \item activation = 'relu'; 
        \item learning rate = 'invscaling';
        \item solver = 'lbfgs'; 
        \item hidden layers sizes = (39, 33, 3); 
        \item \textit{Lag  Association ARIMA} = 7; 
        \item \textit{Lag Association Error} = 18; 
        \item \textit{Forecast Association Error} = 8.
    \end{itemize}
\end{itemize}

\subsubsection{BJL Solar}
In this series the last 64 points are used as test data. The first 256 for training. For this series, seasonal information was used, resulting in a SARIMAX model (Seasonal AutoRegressive Integrated Moving Average with eXogenous regressors model). This result was obtained from 30 individuals and 5 training periods of the evolutionary algorithm used. For this series the final models found by the algorithms are described below:
\begin{itemize}
\item ARIMA: P=1, D=1, Q=1; 
\item Seasonal ARIMA: P=1, D=0, Q=1, s= Monthly 
\item MLP For Error Modeling
    \begin{itemize}
        \item activation='relu'; 
        \item learning rate = 'adaptive';
        \item solver='adam';
        \item hidden layers sizes = (111, 24, 5); 
        \item \textit{Lag Error} = 4.
    \end{itemize}
\item MLP For Nonlinear Association Function
    \begin{itemize}
        \item activation = 'identity'; 
        \item learning rate = 'invscaling';
        \item solver = 'lbfgs'; 
        \item hidden layers sizes = (45, 43, 7); 
        \item \textit{Lag  Association ARIMA} = 8; 
        \item \textit{Lag Association Error} = 5; 
        \item \textit{Forecast Association Error} = 18.
    \end{itemize}
\end{itemize}


\section{Discussion}
\label{sec:discussao}
The results obtained show that the use of AI techniques is promising to assist in the estimation of the energy production of photovoltaic plants, both for the estimation of the capacity factor and for the generation of a new energy production indicator. Below are some ideas for future work.

In the methodology explained in the section \ref{sec:cov-cor} it is possible to add the use of a search algorithm, in order to return weights to each variable and its covariance and correlation.

In the section \ref{sec:espacial_nordeste} a more elaborate interpolation for INMET data can be suggested, but the fact that INMET collection stations are strategically chosen depending on the characteristics of each microclimate in the Northeast, serves as an argument for not need for a different interpolation.

Regarding the results presented regarding the section \ref{sec:otmizacao_comite}, a better way to evaluate would be to make a statistical comparison, based on several executions of both the optimization algorithm of committees used, and of the standard initializations of the regressors, preferably defining parameters random for these.

\bibliographystyle{IEEEtran}
\bibliography{refs}

\begin{thebibliography}{10}
\providecommand{\url}[1]{#1}
\csname url@samestyle\endcsname
\providecommand{\newblock}{\relax}
\providecommand{\bibinfo}[2]{#2}
\providecommand{\BIBentrySTDinterwordspacing}{\spaceskip=0pt\relax}
\providecommand{\BIBentryALTinterwordstretchfactor}{4}
\providecommand{\BIBentryALTinterwordspacing}{\spaceskip=\fontdimen2\font plus
\BIBentryALTinterwordstretchfactor\fontdimen3\font minus
  \fontdimen4\font\relax}
\providecommand{\BIBforeignlanguage}[2]{{%
\expandafter\ifx\csname l@#1\endcsname\relax
\typeout{** WARNING: IEEEtran.bst: No hyphenation pattern has been}%
\typeout{** loaded for the language `#1'. Using the pattern for}%
\typeout{** the default language instead.}%
\else
\language=\csname l@#1\endcsname
\fi
#2}}
\providecommand{\BIBdecl}{\relax}
\BIBdecl

\bibitem{chin2015cell}
V.~J. Chin, Z.~Salam, and K.~Ishaque, ``Cell modelling and model parameters
  estimation techniques for photovoltaic simulator application: A review,''
  \emph{Applied Energy}, vol. 154, pp. 500--519, 2015.

\bibitem{jordehi2016parameter}
A.~R. Jordehi, ``Parameter estimation of solar photovoltaic (pv) cells: A
  review,'' \emph{Renewable and Sustainable Energy Reviews}, vol.~61, pp.
  354--371, 2016.

\bibitem{de2017performance}
C.~L. de~Azevedo~Dias, D.~A.~C. Branco, M.~C. Arouca, and L.~F.~L. Legey,
  ``Performance estimation of photovoltaic technologies in brazil,''
  \emph{Renewable Energy}, vol. 114, pp. 367--375, 2017.

\bibitem{mueller2009cm}
R.~Mueller, C.~Matsoukas, A.~Gratzki, H.~Behr, and R.~Hollmann, ``The cm-saf
  operational scheme for the satellite based retrieval of solar surface
  irradiance—a lut based eigenvector hybrid approach,'' \emph{Remote Sensing
  of Environment}, vol. 113, no.~5, pp. 1012--1024, 2009.

\bibitem{huld2012new}
T.~Huld, R.~M{\"u}ller, and A.~Gambardella, ``A new solar radiation database
  for estimating pv performance in europe and africa,'' \emph{Solar Energy},
  vol.~86, no.~6, pp. 1803--1815, 2012.

\bibitem{amillo2014new}
A.~Amillo, T.~Huld, and R.~M{\"u}ller, ``A new database of global and direct
  solar radiation using the eastern meteosat satellite, models and
  validation,'' \emph{Remote sensing}, vol.~6, no.~9, pp. 8165--8189, 2014.

\bibitem{habte2017evaluation}
A.~Habte, M.~Sengupta, and A.~Lopez, ``Evaluation of the national solar
  radiation database (nsrdb): 1998-2015,'' National Renewable Energy
  Lab.(NREL), Golden, CO (United States), Tech. Rep., 2017.

\bibitem{voyant2017machine}
C.~Voyant, G.~Notton, S.~Kalogirou, M.-L. Nivet, C.~Paoli, F.~Motte, and
  A.~Fouilloy, ``Machine learning methods for solar radiation forecasting: A
  review,'' \emph{Renewable Energy}, vol. 105, pp. 569--582, 2017.

\bibitem{wolff2016statistical}
B.~Wolff, E.~Lorenz, and O.~Kramer, ``Statistical learning for short-term
  photovoltaic power predictions,'' in \emph{Computational
  sustainability}.\hskip 1em plus 0.5em minus 0.4em\relax Springer, 2016, pp.
  31--45.

\bibitem{li2016hierarchical}
Z.~Li, S.~Rahman, R.~Vega, and B.~Dong, ``A hierarchical approach using machine
  learning methods in solar photovoltaic energy production forecasting,''
  \emph{Energies}, vol.~9, no.~1, p.~55, 2016.

\bibitem{khan2014optimal}
G.~Khan and S.~Rathi, ``Optimal site selection for solar pv power plant in an
  indian state using geographical information system (gis),''
  \emph{International Journal of Emerging Engineering Research and Technology},
  vol.~2, no.~7, pp. 260--266, 2014.

\bibitem{fernandez2015site}
L.~A. Fernandez-Jimenez, M.~Mendoza-Villena, P.~Zorzano-Santamaria,
  E.~Garcia-Garrido, P.~Lara-Santillan, E.~Zorzano-Alba, and A.~Falces, ``Site
  selection for new pv power plants based on their observability,''
  \emph{Renewable energy}, vol.~78, pp. 7--15, 2015.

\bibitem{carrion2008electricity}
J.~A. Carri{\'o}n, A.~E. Estrella, F.~A. Dols, and A.~R. Ridao, ``The
  electricity production capacity of photovoltaic power plants and the
  selection of solar energy sites in andalusia (spain),'' \emph{Renewable
  Energy}, vol.~33, no.~4, pp. 545--552, 2008.

\bibitem{boran2010multi}
F.~Boran, T.~Menlik, and K.~Boran, ``Multi-criteria axiomatic design approach
  to evaluate sites for grid-connected photovoltaic power plants: A case study
  in turkey,'' \emph{Energy Sources, Part B: Economics, Planning, and Policy},
  vol.~5, no.~3, pp. 290--300, 2010.

\bibitem{ONS}
``Dados da geração solar fotovoltaica no sin,''
  \url{http://www.ons.org.br/Paginas/resultados-da-operacao/boletim-geracao-solar.aspx},
  accessed: 2019-09-03.

\bibitem{ANEEL}
``Matriz de energia elétrica,''
  \url{http://www2.aneel.gov.br/aplicacoes/capacidadebrasil/OperacaoCapacidadeBrasil.cfm},
  accessed: 2019-09-03.

\bibitem{pereira2017atlas}
E.~Pereira, F.~Martins, A.~Gon{\c{c}}alves, R.~Costa, F.~Lima, R.~R{\"u}ther,
  S.~Abreu, G.~Tiepolo, S.~Pereira, and J.~Souza, ``Atlas brasileiro de energia
  solar--2{\textordfeminine} edi{\c{c}}{\~a}o, s{\~a}o jos{\'e} dos campos,''
  2017.

\bibitem{LABREN}
``Atlas brasileiro de energia solar,''
  \url{http://labren.ccst.inpe.br/atlas_2017.html}, accessed: 2019-09-04.

\bibitem{martins2005base}
F.~R. Martins, E.~B. Pereira, C.~Yamashita, S.~V. Pereira, and
  S.~Mantelli~Neto, ``Base de dados clim{\'a}tico-ambientais aplicados ao setor
  energ{\'e}tico-projeto sonda,'' \emph{Proc. of XII Simp{\'o}sio Brasileiro de
  Sensoriamento Remoto, INPE, Sao Jos{\'e} dos Campos, Brazil}, 2005.

\bibitem{martins2007mapeamento}
F.~R. Martins, E.~B. Pereira, R.~A. Guarnieri, S.~A. Silva, C.~S. Yamashita,
  R.~C. Chagas, S.~L. Abreu, and S.~Colle, ``Mapeamento dos recursos de energia
  solar no brasil uti-lizando modelo de transfer{\^e}ncia radiativa
  brasil-sr,'' in \emph{Anais do I Congresso Brasileiro de Energia Solar},
  2007, pp. 8--10.

\bibitem{INMET}
``Instituto nacional de meteorologia,'' \url{http://www.inmet.gov.br/portal/},
  accessed: 2019-10-02.

\bibitem{PVGIS}
``Photovoltaic geographical information system,''
  \url{https://re.jrc.ec.europa.eu/pvgis.html}, accessed: 2019-09-04.

\bibitem{muller2015sarah}
R.~M{\"u}ller, J.~Trentmann, C.~Tr{\"a}ger-Chatterjee, and U.~Pfeifroth,
  ``Sarah-a new homogeneous climate data record of surface radiation,'' in
  \emph{EGU General Assembly Conference Abstracts}, vol.~17, 2015.

\bibitem{pvgis_web}
``Interface for accessing pvgis data and calculations,''
  \url{https://re.jrc.ec.europa.eu/pvg_static/web_service.html}, accessed:
  2019-10-02.

\bibitem{PVGIS_methods}
``Overview of pvgis data sources and calculation methods,''
  \url{https://re.jrc.ec.europa.eu/pvg_static/methods.html}, accessed:
  2019-10-11.

\bibitem{PVGIS_bases}
``Other free solar radiation data and pv tools,''
  \url{https://re.jrc.ec.europa.eu/pvg_static/data_sources.html}, accessed:
  2019-10-11.

\bibitem{stein2012interpolation}
M.~L. Stein, \emph{Interpolation of spatial data: some theory for
  kriging}.\hskip 1em plus 0.5em minus 0.4em\relax Springer Science \& Business
  Media, 2012.

\bibitem{pykrige}
\BIBentryALTinterwordspacing
P.~developers, \emph{PyKrige Documentation}, 2019, release 1.4.1. [Online].
  Available: \url{https://pykrige.readthedocs.io/en/latest/index.htm}
\BIBentrySTDinterwordspacing

\bibitem{oliver2014tutorial}
M.~Oliver and R.~Webster, ``A tutorial guide to geostatistics: Computing and
  modelling variograms and kriging,'' \emph{Catena}, vol. 113, pp. 56--69,
  2014.

\bibitem{svetnik2003random}
V.~Svetnik, A.~Liaw, C.~Tong, J.~C. Culberson, R.~P. Sheridan, and B.~P.
  Feuston, ``Random forest: a classification and regression tool for compound
  classification and qsar modeling,'' \emph{Journal of chemical information and
  computer sciences}, vol.~43, no.~6, pp. 1947--1958, 2003.

\bibitem{drucker1997support}
H.~Drucker, C.~J. Burges, L.~Kaufman, A.~J. Smola, and V.~Vapnik, ``Support
  vector regression machines,'' in \emph{Advances in neural information
  processing systems}, 1997, pp. 155--161.

\bibitem{solomatine2004adaboost}
D.~P. Solomatine and D.~L. Shrestha, ``Adaboost. rt: a boosting algorithm for
  regression problems,'' in \emph{2004 IEEE International Joint Conference on
  Neural Networks (IEEE Cat. No. 04CH37541)}, vol.~2.\hskip 1em plus 0.5em
  minus 0.4em\relax IEEE, 2004, pp. 1163--1168.

\bibitem{breiman1996bagging}
L.~Breiman, ``Bagging predictors,'' \emph{Machine learning}, vol.~24, no.~2,
  pp. 123--140, 1996.

\bibitem{friedman2002stochastic}
J.~H. Friedman, ``Stochastic gradient boosting,'' \emph{Computational
  statistics \& data analysis}, vol.~38, no.~4, pp. 367--378, 2002.

\bibitem{fischler1981random}
M.~A. Fischler and R.~C. Bolles, ``Random sample consensus: a paradigm for
  model fitting with applications to image analysis and automated
  cartography,'' \emph{Communications of the ACM}, vol.~24, no.~6, pp.
  381--395, 1981.

\bibitem{crammer2006online}
K.~Crammer, O.~Dekel, J.~Keshet, S.~Shalev-Shwartz, and Y.~Singer, ``Online
  passive-aggressive algorithms,'' \emph{Journal of Machine Learning Research},
  vol.~7, no. Mar, pp. 551--585, 2006.

\bibitem{scikit-learn}
F.~Pedregosa, G.~Varoquaux, A.~Gramfort, V.~Michel, B.~Thirion, O.~Grisel,
  M.~Blondel, P.~Prettenhofer, R.~Weiss, V.~Dubourg, J.~Vanderplas, A.~Passos,
  D.~Cournapeau, M.~Brucher, M.~Perrot, and E.~Duchesnay, ``Scikit-learn:
  Machine learning in {P}ython,'' \emph{Journal of Machine Learning Research},
  vol.~12, pp. 2825--2830, 2011.

\bibitem{sun2017correlation}
Y.~Sun, F.~Wang, B.~Wang, Q.~Chen, N.~Engerer, and Z.~Mi, ``Correlation feature
  selection and mutual information theory based quantitative research on
  meteorological impact factors of module temperature for solar photovoltaic
  systems,'' \emph{Energies}, vol.~10, no.~1, p.~7, 2017.

\bibitem{kazem2015effect}
H.~A. Kazem and M.~T. Chaichan, ``Effect of humidity on photovoltaic
  performance based on experimental study,'' \emph{International Journal of
  Applied Engineering Research (IJAER)}, vol.~10, no.~23, pp. 43\,572--43\,577,
  2015.

\bibitem{hailegnaw2015rain}
B.~Hailegnaw, S.~Kirmayer, E.~Edri, G.~Hodes, and D.~Cahen, ``Rain on
  methylammonium lead iodide based perovskites: possible environmental effects
  of perovskite solar cells,'' \emph{The journal of physical chemistry
  letters}, vol.~6, no.~9, pp. 1543--1547, 2015.

\bibitem{zhang2003time}
G.~P. Zhang, ``Time series forecasting using a hybrid arima and neural network
  model,'' \emph{Neurocomputing}, vol.~50, pp. 159--175, 2003.

\bibitem{khashei2010artificial}
M.~Khashei and M.~Bijari, ``An artificial neural network (p, d, q) model for
  timeseries forecasting,'' \emph{Expert Systems with applications}, vol.~37,
  no.~1, pp. 479--489, 2010.

\bibitem{babu2014moving}
C.~N. Babu and B.~E. Reddy, ``A moving-average filter based hybrid arima--ann
  model for forecasting time series data,'' \emph{Applied Soft Computing},
  vol.~23, pp. 27--38, 2014.

\bibitem{de2014hybrid}
J.~F.~L. de~Oliveira and T.~B. Ludermir, ``A hybrid evolutionary system for
  parameter optimization and lag selection in time series forecasting,'' in
  \emph{2014 Brazilian Conference on Intelligent Systems}.\hskip 1em plus 0.5em
  minus 0.4em\relax IEEE, 2014, pp. 73--78.

\bibitem{de2016hybrid}
J.~F. de~Oliveira and T.~B. Ludermir, ``A hybrid evolutionary decomposition
  system for time series forecasting,'' \emph{Neurocomputing}, vol. 180, pp.
  27--34, 2016.

\bibitem{domingos2019intelligent}
S.~d.~O. Domingos, J.~F. de~Oliveira, and P.~S. de~Mattos~Neto, ``An
  intelligent hybridization of arima with machine learning models for time
  series forecasting,'' \emph{Knowledge-Based Systems}, vol. 175, pp. 72--86,
  2019.

\end{thebibliography}

\end{document}